\documentclass[11pt]{article}
\usepackage{mathrsfs}
\usepackage{natbib,graphicx,setspace,lscape,longtable}
\usepackage{mathrsfs,amsmath,amsthm,amssymb,color}
\usepackage{natbib,epsfig,graphicx,pdfpages}
\usepackage{rotating}
\usepackage{subfigure}
\usepackage{graphicx}
\usepackage{url}

\bibpunct{(}{)}{;}{a}{,}{,}

\newtheorem{theorem}{Theorem}
\newtheorem{lemma}{Lemma}
\newtheorem{proposition}{Proposition}
\newtheorem{remark}{Remark}
\usepackage{amsthm}

\newtheorem{assumption}{{Assumption}}
\def\beq{\begin{equation}}
\def\eeq{\end{equation}}
\def\beqr{\begin{eqnarray}}
\def\eeqr{\end{eqnarray}}
\def\beqrs{\begin{eqnarray*}}
\def\eeqrs{\end{eqnarray*}}
\def\bet{\begin{theorem}}
\def\eet{\end{theorem}}
\def\bel{\begin{lemma}}
\def\eel{\end{lemma}}
\def\bep{\begin{proposition}}
\def\eep{\end{proposition}}
\def\bg{\begin{figure}[tbph]\begin{center}}
\def\eg{\end{center}\end{figure}}

\def\bc{\begin{center}}
\def\ec{\end{center}}

\def\wt{\widetilde}
\def\wh{\widehat}

\def\diag{\mbox{diag}}

\numberwithin{equation}{section}

\newcommand{\Cov}{\textnormal{Cov}}

\newcommand{\bA}{{\mathbf A}}

\newcommand{\bH}{{\mathbf H}}
\newcommand{\bI}{{\mathbf I}}

\newcommand{\bL}{{\mathbf L}}
\newcommand{\bM}{{\mathbf M}}

\newcommand{\bP}{{\mathbf P}}

\newcommand{\bV}{{\mathbf V}}

\newcommand{\ba}{{\mathbf a}}

\newcommand{\bd}{{\mathbf d}}
\newcommand{\be}{{\mathbf e}}
\newcommand{\bff}{{\mathbf f}}

\newcommand{\bs}{{\mathbf s}}
\newcommand{\bu}{{\mathbf u}}

\newcommand{\bx}{{\mathbf x}}
\newcommand{\by}{{\mathbf y}}
\newcommand{\bz}{{\mathbf z}}
\newcommand{\balpha} {\boldsymbol{\alpha}}
\newcommand{\bbeta}  {\boldsymbol{\beta}}
\newcommand{\bfeta}  {\boldsymbol{\eta}}

\newcommand{\bSigma}{\boldsymbol{\Sigma}}

\newcommand{\bgamma}{\boldsymbol{\gamma}}

\newcommand{\bve}{\mbox{\boldmath$\varepsilon$}}

\newcommand{\bTheta} {\boldsymbol{\Theta}}
\newcommand{\bPhi} {\boldsymbol{\Phi}}

\newcommand{\btheta} {\boldsymbol{\theta}}
\newcommand{\bxi} {\boldsymbol{\xi}}
\newcommand{\bmu} {\boldsymbol{\mu}}

\newcommand{\bD}{{\mathbf D}}
\newcommand{\bzero}{{\mathbf 0}}

\newcommand{\ve}{{\varepsilon}}
\renewcommand{\epsilon}{{\ve}}
\renewcommand{\hat}{\widehat}
\newcommand{\T}{{\rm T}}
\def\wt{\widetilde}

\def\BKA{{\sl Biometrika }}
\def\JASA{{\sl Journal of the American Statistical Association }}

\begin{document}

\title{A Structural-Factor Approach to Modeling High-Dimensional Time Series and Space-Time Data}

\author{
Zhaoxing Gao and Ruey S Tsay \\
Booth School of Business, University of Chicago
}


\maketitle

\begin{abstract}
This paper considers a structural-factor approach to modeling high-dimensional time series 
and space-time data by decomposing individual series into trend, seasonal, and irregular components. For ease in analyzing 
many time series, we employ a time polynomial for the trend, a linear combination of trigonometric series for the seasonal component, and 
a new factor model for the irregular components. 
The new factor model simplifies the modeling process and achieves parsimony in parameterization. 
We propose a Bayesian Information Criterion (BIC) to consistently select the order of the polynomial trend and the number of trigonometric functions, and use 
a test statistic to determine the number of common factors. 
The convergence rates for the  estimators of 
 the trend and seasonal components 
and the limiting distribution of the test statistic
are established under the setting that the number of time series tends to infinity with the sample size, but at a slower rate. We study 
 the finite-sample performance of the proposed analysis via simulation, 
and analyze two real examples. 
The first example considers modeling weekly PM$_{2.5}$ data of 
15 monitoring stations in the southern region of Taiwan and the second 
example consists of monthly value-weighted returns of 12 industrial portfolios. 
\end{abstract}


\newpage

\section{Introduction}
The availability of high-dimensional time series and space-time data under the current 
big-data environment 
creates new challenges in time series modeling, and analysis of such data 
has emerged as an important and active research area in many 
scientific fields, including  engineering, environmental studies, and statistics. 
In theory, the vector autoregressive moving-average (VARMA) models can be used, 
but their applications often encounter the 
difficulties of over-parametrization and lack of identifiability, 
especially when the dimension is high.  
Over-parametrization is likely to occur when one uses unrestricted VARMA models,  
and it is well-known that exchangeable models exist in VARMA specification.
See, for instance, \cite{TiaoTsay_1989}, L\"{u}tkepohl (2006),  Tsay (2014), and the 
references therein. Various methods have been developed to overcome the identifiability 
issues and to reduce the number of parameters of VARMA models. 
For example,  Chapter 4 
of \cite{Tsay_2014}, and the references therein, discussed various canonical structures of a VARMA model. \cite{Davis2012} studied the vector autoregressive (VAR) model with sparse coefficient matrices based on partial spectral coherence. 
The Lasso regularization has also been applied to VAR models to reduce 
the number of parameters;  see \cite{ShojaieMichailidis_2010} and  \cite{SongBickel_2011}, among others. \cite{GuoWangYao_2014} considered banded VAR models and estimated the coefficient matrices by a componentwise least squares method. For dimension reduction, popular methods include the canonical correlation analysis (CCA) of \cite{BoxTiao_1977}, the principle component analysis (PCA) of \cite{StockWatson_2002}, and 
the scalar component analysis of \cite{TiaoTsay_1989}. 
An alternative approach to analyzing high-dimensional time series is to employ factor models; see, for instance, \cite{BaiNg_Econometrica_2002}, \cite{StockWatson_2005}, \cite{panyao2008}, \cite{LamYaoBathia_Biometrika_2011}, \cite{lamyao2012} and \cite{changguoyao2015}. In fact, the idea of latent factors driving common 
behavior in multiple time series can be dated back, at least, to Nerlove (1964). 
Most of the factor models considered in the literature assume weak stationarity of the 
underlying  time series and employ latent factors to describe the overall temporal 
dependence of the data.

Empirical time series often exhibit complex patterns, which may include trend and seasonal components. For example, the hourly measurements of fine particulate matter 
(PM$_{2.5}$) 
at different monitoring stations in a given region show not only an annual cycle but also certain 
diurnal pattern possibly caused by wind direction, wind speed, humidity, temperature, 
and human activities. The measurements may also exhibit some trending behavior due to increased urbanization, as some studies state that the urbanization level plays a positive role in promoting carbon emission, which is a major component of particulate matters. 
See, for example, \cite{zhang2015}. In the time series literature, 
structural models consisting of trend, seasonal, and irregular components 
have been proposed 
to analyze univariate series with complex patterns. See, for instance, Harvey (1989). 
As a matter of fact, a range of trend and periodic analyses for environmental and economic time series have appeared in the literature; see \cite{wallis1978}, \cite{plosser1979}, Barsky and Miron (1989), Harvey and Koopman (1993), Chang et al. (2009), De Livera et al. (2011), among others. 
However, none of those methods can be applied (or have been extended) to model jointly 
high-dimensional time series. Several methods have also been 
developed in the spatio-temporal 
literature to explore the spatial and temporal dependence of the data. 
See, for instance,  Yu et al. (2008), Lee and Yu (2010),
Lin and Lee (2010), Kelejian and Prucha (2010), Su (2012), and Gao et al. (2019), among others. However, most of the available methods require specification of a spatial weight matrix or an appropriate ordering of the locations. For a given application, 
the choices  may not be obvious, and the resulting spatial autoregressive model may fail to accommodate adequately 
the dependent structure among different locations.

\begin{figure}
\begin{center}
\subfigure[]{\includegraphics[width=3.0in,height=2.7in]{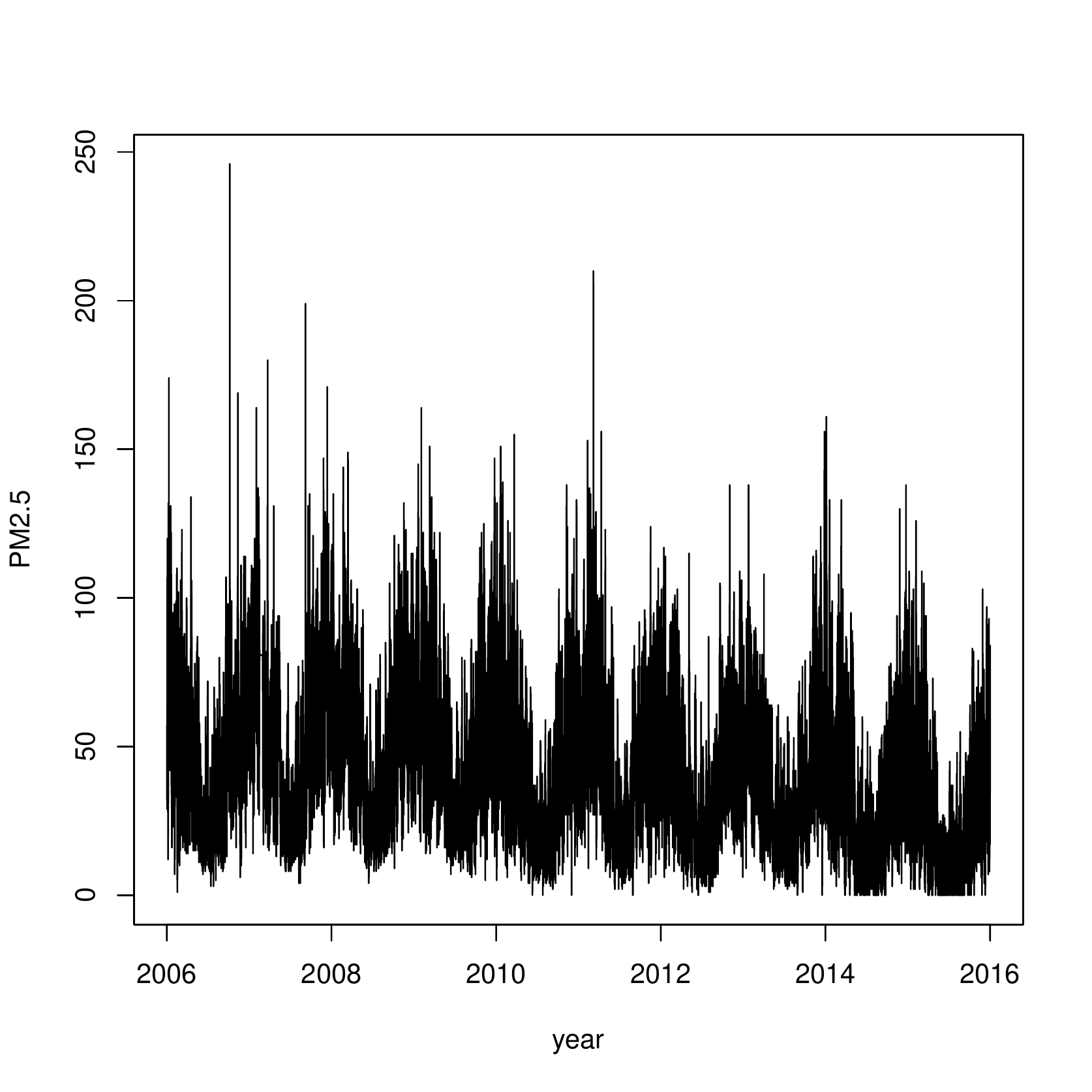}}
\subfigure[]{\includegraphics[width=3.0in,height=2.7in]{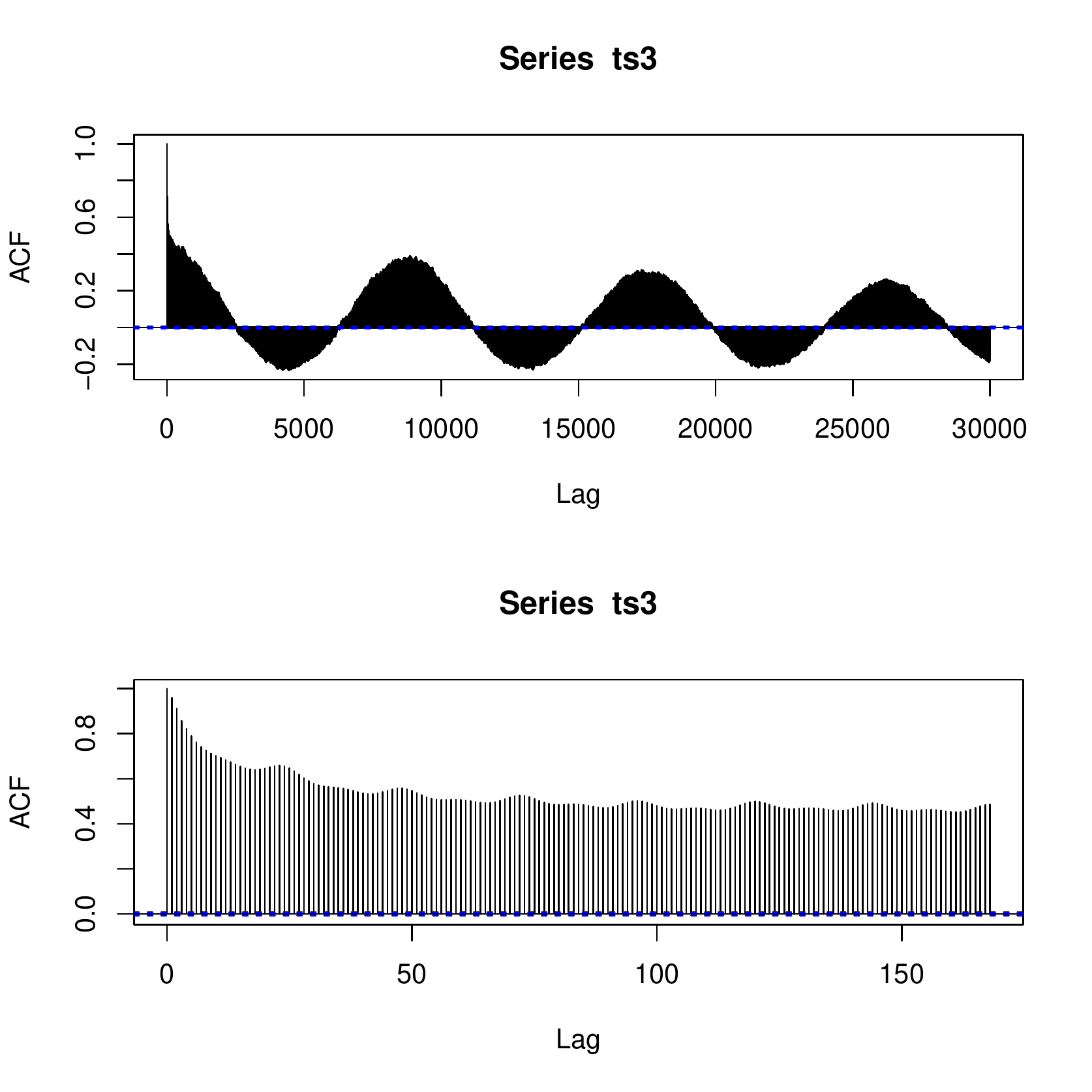}}
\end{center}
\caption{(a) Time plot of hourly PM$_{2.5}$ at a monitoring station in Taiwan from 
January 1, 2006 to December 31, 2015 for 87,600 observations; (b) Sample autocorrelation functions of the hourly time series of (a).}\label{ts3}
\end{figure}

The goal of this paper is to combine the 
structural models with latent factors for analysis of high-dimensional time series and 
space-time data. By focusing on 
common factors for the irregular components rather than on the observed series directly, 
the proposed models can be more effective in identifying the 
number of common factors and 
can provide further insight in understanding the latent structure of the data. For instance, 
the identified common latent factors of the irregular components are free of the effects of 
trend or seasonality. Furthermore, the combined approach can leverage the advantages of 
structural specification and factor models. This is particularly relevant in analyzing high-dimensional 
and high-frequency data for which the common patterns can be complex and 
the observed data are typically non-Gaussian. Consider, for instance, 
the hourly measurements of PM$_{2.5}$ at a monitoring station. 
Such a series is often not normally distributed and exhibits 
annual, weekly, and diurnal patterns in addition to the local trending behavior. 
Figure~\ref{ts3}(a) 
shows the time plot of hourly PM$_{2.5}$ measurements 
at a monitoring station in Taiwan from 
January 1, 2006 to December 31, 2015 for 87,600 observations 
(the data for February 29 were 
removed for simplicity). The traditional test statistics for skewness and excess kurtosis 
assume the values 99.5 and 56.9, respectively, with $p$-values close to zero. 
Figure~\ref{ts3}(b) shows the sample autocorrelation functions 
(SACF) 
of the series. The upper plot contains 30,000 lags of SACF and it exhibits clearly an 
annual cycle with frequency $\omega = 2\pi/8760$ = $7.17\times 10^{-4}$. The lower plot 
shows the first 200 lags of SACF and it also exhibits clearly an diurnal pattern with 
frequency $\omega_1 = 2\pi/24$ = 0.262.  Thus, the cyclical patterns of the series are  
complex. Furthermore, the periodicity of 8760 is sufficiently large, making it hard to 
fit a seasonal ARMA model to the data. 
Finally, consider the situation in which many such 
PM$_{2.5}$ time series are available. It would not be simple to identify the overall 
latent factors among the series. This type of problem is not unique to environment studies. 
As a matter of fact,  big data indexed by two or more  indexes, such as the 
PM$_{2.5}$ in space and time, are routinely collected nowadays in many scientific fields. 
For example, in economics, monthly consumer price index (CPI) are collected for every member 
country of the European Union, and the state unemployment rates are compiled for the 50 states 
in the US. In finance, asset returns of many industrial portfolios or exchange-traded funds (ETF) are available daily. 
These data also exhibit complex cyclical (or seasonal) patterns. 
See \cite{changmiller2009} and \cite{changpinegar1989} for details. 
In addition, researchers also found that  a large number of economic variables can be
modeled by a small number of factors, which provides a
convenient way to study the aggregate implications of microeconomic behavior,
as shown in \cite{forni1997}. 
Examples such as these mentioned above motivate us to consider the proposed approach. 

In applications, we often care about the general direction of the observed data over time. 
Therefore, for simplicity, we use a deterministic polynomial for the trend component and 
 a linear combination of  trigonometric functions for the seasonal component. 
This simplifying assumption is used mainly to overcome the difficulty in handling 
the high periodicity as that shown in the hourly PM$_{2.5}$ example. 
For the irregular components, we 
employ a new factor model to reduce the number of parameters and to describe the 
common stochastic characteristics of the data. The proposed factor model differs from the 
factor models commonly used in the literature, as we seek a nonsingular linear  transformation that separates white noise series from the dynamically dependent ones. 
From a dimension reduction point of view, we treat the polynomial and trigonometric basis functions as the factors for the trend and seasonal components, respectively, and the latent 
stochastic factors for the irregular parts. The latent factors are linear combinations of the 
irregular components and are estimated by a canonical correlation analysis. 
The number of white noise series is determined by a test statistic. 
We propose a Bayesian Information Criterion (BIC) to consistently select the order of the polynomial trend and the number of the trigonometric functions. 
We also derive the convergence rates for the coefficient estimators  of the trend and seasonal components and the limiting properties of the test statistic 
under the setting that the dimension of the time series tends to infinity with the sample size, but at a slower rate. Finally, we use simulation and 
real data examples to assess the performance of the proposed procedure.

The rest of the paper is organized as follows. We specify the proposed methodology in Section 2 
with special attention being paid to the new factor model for the irregular components. 
The differences between the new model and the commonly used factor models in the literature 
are given. In Section 3, we study 
the theoretical properties of the proposed model. 
Section 4 reports the results of simulation studies and empirical applications of 
two examples. 
Section 5 provides concluding remarks. All technical proofs are relegated to an online supplement. We use the following notation, for a $p\times 1$ vector
$\bu=(u_1,..., u_p)^\T,$  $||\bu||_2 = (\sum_{i=1}^{p} u_i^2)^{1/2} $
is the Euclidean norm, and $\bI_p$ denotes the $p\times p$ identity matrix. For a matrix $\bH=(h_{ij})$, $|\bH|_\infty=\max_{i,j}|h_{ij}|$,  $\|\bH
\|_2=\sqrt{\lambda_{\max} (\bH^\T \bH ) }$ is the operator norm, where
$\lambda_{\max} (\cdot) $ denotes for the largest eigenvalue of a matrix, and $\|\bH\|_{\min}$ is the square root of the minimum eigenvalue of $\bH^\T\bH$. The superscript $\T$ denotes 
the transpose of a vector or matrix. Finally, we use the notation $a\asymp b$ to denote $a=O(b)$ and $b=O(a)$.

\section{Methodology}
\subsection{The Setting and Method}
Let $\by_t=(y_{1t},...,y_{pt})^{\T}$ be a $p$-dimensional time series or an 
observation from $p$ spatial locations at time $t$. 
We assume that 
\begin{equation}\label{decom}
\by_t=\bmu_t+\bs_t+\bfeta_t,
\end{equation}
where $\bmu_t$, $\bs_t$ and $\bfeta_t$ denote, respectively, the trend, seasonal and irregular components with $\bmu_t=(\mu_{1t},...,\mu_{pt})^{\T}$, $\bs_t=(s_{1t},...,s_{pt})^{\T}$, and $\bfeta_t=(\eta_{1t},...,\eta_{pt})^\T$. 
For each $1\leq i\leq p$, we assume
\begin{equation}\label{trend-season}
\mu_{it}=\alpha_{i0}+\alpha_{i1}t+...+\alpha_{id_0}t^{d_0}\quad\text{and}\quad s_{it}=\sum_{j=1}^{k_0}[\beta_{ij}\cos(\rho_j t)+\gamma_{ij}\sin(\rho_j t)],
\end{equation}
where $\rho_j=2\pi j/s$ with $s$ being a known periodicity, and $d_0$ and $k_0$ are nonnegative  integers. The irregular component $\bfeta_t$ is weakly stationary with 
$E(\bfeta_t) = \bzero$ and can be written as
\begin{equation}\label{cca-transform}
\bfeta_t = \wt\bL\left[\begin{array}{c}\bff_t \\ \bve_t\end{array}\right]=\wt\bL_1\bff_t+\wt\bL_2\bve_t,
\end{equation}
where $\wt\bL=(\wt\bL_1,\wt\bL_2)$ is a $p\times p$  nonsingular (real-valued) 
matrix, $\bff_t = (f_{1t},\ldots,
f_{rt})^\T$ and $\bve_t = (\varepsilon_{1t},\ldots,\varepsilon_{vt})^\T$ with $r$ and $v$ being 
nonnegative integers such that $r+v = p$. 
Let $\bV^\T=\wt\bL^{-1}$ and $\bV=(\bV_1,\bV_2)$ with $\bV_1\in R^{p\times r}$ and $\bV_2\in R^{p\times v}$. Model (\ref{cca-transform}) is a transformation model employed in  \cite{TiaoTsay_1989}. Specifically, 
there exists a nonsingular transformation matrix $\bV$ such that $\bV_1^\T\bfeta_t=\bff_t$ and $\bV_2^\T\bfeta_t=\bve_t$ with dimensions $r$ and $v$, respectively. 

In Equation (\ref{cca-transform}), we assume that (a) $\bve_t$ is a $v$-dimensional scalar component process of 
order (0,0) if $v > 0$, that is, 
Cov($\bve_t,\bfeta_{t-j}$) = $\bzero$ for $j > 0$, and 
(b) no linear combination of $\bff_t$ is a scalar component of order (0,0) if $r > 0$. 
Assumption (b) is obvious; otherwise $v$ can be increased. 
For the formal definition of a general scalar component of order $(p_1,q_1)$ with $p_1,q_1\geq 0$, readers are 
referred to \cite{TiaoTsay_1989}.  Assumption (a) is equivalent to $\bve_t$ 
being a white noise under the traditional factor models for which $\bff_t$ and 
$\bve_t$ are assumed to be independent.

Any finite-order VARMA process $\bfeta_t$ can always be written in Equation (\ref{cca-transform}) 
via canonical correlation analysis (CCA) between two constructed vectors of 
$\bfeta_t$ and its lagged variables. 
See \cite{TiaoTsay_1989}. Also, readers are referred to Chapter 12 of \cite{anderson1958} for an  introduction of CCA between two random vectors. 
 Under Equation (\ref{cca-transform}), the dynamic dependence of $\bfeta_t$  
is driven by $\bff_t$ if $r > 0$. In this sense, $\bff_t$ indeed consists of the common factors of 
$\bfeta_t$. 

In contrast to Equation (\ref{cca-transform}), the most commonly used 
factor model in the literature is 
\begin{equation}\label{factor}
\bfeta_t=\bA\bx_t+\bve_t,
\end{equation}
where $\bx_t\in R^{r}$ is a latent factor process, $\bA\in R^{p\times r}$ is an unknown factor loading matrix,  $\bve_t=(\varepsilon_{1t},...,\varepsilon_{pt})^\T$ 
is a serially uncorrelated process with mean {$\bf 0$} and covariance matrix $\bSigma_\ve$, 
and $\bx_t$ and $\bve_t$ are independent. See, 
for instance, \cite{lamyao2012}. A major difference between Equations (\ref{cca-transform}) 
and (\ref{factor}) is that the right side of Equation (\ref{cca-transform}) has 
$p$ random errors 
whereas that of Equation (\ref{factor}) has $r+p$ random errors. 
In this paper, we refer to the random errors, which consist of serially uncorrelated random 
variables with mean zero and finite covariance matrix, as innovations to a time series. 
Consequently, the sample covariance matrix of the innovation $\bve_t$ of Equation (\ref{factor}) 
is always singular if $r > 0$. 
On the other hand, the sample covariance matrix of the innovation $\bve_t$ of 
Equation (\ref{cca-transform}) is positive-definite provided that the sample size is sufficiently large. 

Assume that $r > 0$ in Equation (\ref{cca-transform}). We further assume that $\bff_t$ follows 
a vector autoregressive model, VAR($d$), 
\begin{equation}\label{var-d}
 \bff_t = \sum_{i=1}^d \bPhi_i \bff_{t-i} + \bu_t,
 \end{equation}
 where $\bu_t$ is a $r$-dimensional innovations with positive diagonal covariance matrix 
 and independent of $\bve_t$. For large $r$, a
 sparse VAR model can be used. Under the VAR assumption in (\ref{var-d}), the process 
 $\bfeta_t$ follows a VAR($d$) model with $\bPhi_i$ forming a non-zero block of the 
 $i$th AR coefficient matrix. 
 For a general $p$-dimensional zero-mean VAR($d$) model, the number of 
parameters is $dp^2+p(p+1)/2$, including those in the error covariance matrix. 
On the other hand, 
 for the proposed model in Equations (\ref{cca-transform}) and (\ref{var-d}), the 
 number of parameters is $p(p-1)/2 + dr^2 + p$, where $p(p-1)/2$ 
 is the number of parameters in the transformation matrix. When $r$ is 
 smaller than $p$, the proposed factor model becomes more parsimonious. For instance, 
 consider the simple case of $p$ = 15, $r = 10$, and $d$ = 1, the reduction in the number of parameters is $345-220$ = 125, which is not a small number. 
In general, the reduction in 
the number of parameters is $d(p^2-r^2)$, which can be substantial if 
either $p$ or $d$ is large and $r$ is small.

In Equation (\ref{cca-transform}),  $(\bL,\bff_t,\bve_t)$ are all latent and any linear transformation of $\bfeta_t$ will not alter the canonical correlation analysis between $\bfeta_t$ and its 
lagged variables. Therefore, we consider the following transformed model
\begin{equation}\label{cca-mod}
\bSigma_{\eta}^{-1/2}\bfeta_t=\bL\left[\begin{array}{c}\bff_t \\ \bve_t\end{array}\right]=\bL_1\bff_t+\bL_2\bve_t,
\end{equation}
where $\bSigma_\eta=\Cov(\bfeta_t)$  and $\bL=(\bL_1,\bL_2)=\bSigma_{\eta}^{-1/2}\wt\bL$ is a $p\times p$  orthonormal matrix. We will see later that this transformation can be done 
via canonical correlation analysis. 

For Model (\ref{cca-mod}), we refer to $\bL_1$ the factor loading matrix 
of $\bff_t$. The matrix $\bL$ and the latent factor $\bff_t$ are not uniquely determined 
in (\ref{cca-mod}). For example, we can replace $(\bL,\bff_t,\bve_t)$ by $(\bL\bH,\bH_1^{-1}\bff_t,\bH_2^{-1}\bve_t)$ for any invertible  diagonal matrix $\bH=\diag(\bH_1,\bH_2)$ and 
(\ref{cca-mod}) still holds. Without loss of generality, we assume $\Cov(\bxi_t)=\bI_p$, 
where $\bxi_t = (\bff_t^{\T},\bve_t^{\T})^{\T}$. 
Under this framework, it follows from (\ref{cca-mod}) that $\bL$ is an orthonormal matrix, i.e. $\bL\bL^\T=\bL^\T\bL=\bI_p$. Note that, since canonical correlations may not be distinct, 
only $\mathcal{M}(\bL_1)$ (and hence $\mathcal{M}(\bL_2)$) can be uniquely determined, where $\mathcal{M}(\bL_1)$ denotes the linear space spanned by the columns of the 
matrix $\bL_1$ and is called the factor loading space.

Given the data $\{\by_t| t=1,...,T\}$, the goal here is to estimate the parameters $\balpha_{i}=(\alpha_{i0},...,\alpha_{id_0})^\T$, $\bbeta_{i}=(\beta_{i1},...,\beta_{ik_0})^\T$,  $\bgamma_{i}=(\gamma_{i1},...,\gamma_{ik_0})^\T$ for $1\leq i\leq p$, $\bL$ and the number 
of common factors $r$, and to recover the factor process $\bff_t$, allowing the dimension $p$ to increase as the sample size 
$T$ increases, where $\balpha_i$, $\bbeta_i$ and $\bgamma_i$ are the coefficients defined in (\ref{trend-season}). 
In practice, $d_0$, $k_0$ and $r$ are also unknown and we propose methods 
to estimate them consistently in Section 2.2.

The proposed methodology is as follows. We first treat $d_0$ and $k_0$ as known integers and let $\btheta_i=(\balpha_i^\T,\bbeta_i^\T,\bgamma_i^\T)^\T$, $\bz_i=(y_{i1},...,y_{iT})^\T$ and $\be_i=(\eta_{i1},...,\eta_{iT})^\T$, where $\bz_i$ and $\be_i$ denote, respectively, the $i$-th component of $\by_t$ and $\bfeta_t$ over time. Define $\bD=(\bd_1,...,\bd_T)^\T$ with $\bd_t=(1,t,...,t^{d_0},\cos(\rho_1 t),...,\cos(\rho_{k_0} t),\sin(\rho_1t),...,\sin(\rho_{k_0}t))^\T$.  Then, (\ref{decom}) can be expressed as
\begin{equation}\label{lse}
\bz_i=\bD\btheta_i+\be_i,\quad i=1, ..., p.
\end{equation}
It follows from (\ref{lse}) that the ordinary least squares estimator (LSE) $\hat{\btheta}_i$ for $\btheta_i$ satisfies
\begin{equation}\label{lse:th}
\hat{\btheta}_i-\btheta_i=(\bD^\T\bD)^{-1}\bD^\T\be_i,\quad i=1, ..., p, 
\end{equation}
and  the associated residuals are $\hat{\bfeta}_t=(\hat{\eta}_{1t},...,\hat{\eta}_{pt})^\T$ with $\hat{\eta}_{it}=y_{it}-{\bd}_t^\T\hat{\btheta}_i$.
Furthermore, the resulting residual sum of squares is
\begin{equation}\label{RSS}
\text{RSS}_i\equiv\text{RSS}_i(k_0,d_0)=\bz_i^\T\{\bI_{T}-\bD(\bD^\T\bD)^{-1}\bD^\T\}\bz_i,\quad i=1, ..., p,
\end{equation}
where $\text{RSS}_i(k_0,d_0)$ is a function of $k_0$ and $d_0$, and we can similarly define $\text{RSS}_i(k,d)$ for any $1\leq k\leq s/2-1$ and $d\geq 0$. Note that $\wh\btheta_i$'s are estimated equation-by-equation. Under the assumption that $\bfeta_t$ is stationary, the estimators are equivalent to those of the generalized least squares method; see Section 2.5.1.1 of \cite{Tsay_2014} for a discussion on the estimation of VAR models.

Turn to the determination of the number of common factors $r$ and the estimation of $\bL$. 
From Equation (\ref{cca-mod}), we have 
\begin{equation}\label{cca-transform-inverse}
\bL^{-1}\bSigma_\eta^{-1/2}\bfeta_t = \bL^\T\bSigma_\eta^{-1/2}\bfeta_t = \left[\begin{array}{c}\bff_t \\ \bve_t\end{array}\right].
\end{equation}
Thus, there are $v$ linear combinations of $\bfeta_t$ 
that are scalar components of order (0,0) 
and we can apply the approach of 
\cite{TiaoTsay_1989} to specify $v$ and, hence, $r = p-v$. 

Let $\bfeta_{t,m} =(\bfeta_{t-1}^\T,\ldots,\bfeta_{t-m}^\T)^\T$ be the vector of 
past $m$ lagged values of $\bfeta_t$, where $m$ is a sufficiently large positive integer. 
Since $\bve_t$ are scalar components of order (0,0), we have Cov($\bve_t$,$\bfeta_{t,m}$) = $\bzero$.
Consequently, there are $v$ zero canonical correlations between $\bfeta_t$ and $\bfeta_{t,m}$. 
Let $\bSigma_{\eta \eta_m} = \Cov(\bfeta_t,\bfeta_{t,m})$ and 
$\bSigma_{\eta_m} = \Cov(\bfeta_{t,m})$. The canonical correlation analysis between $\bfeta_t$ and $\bfeta_{t,m}$ 
is the eigenvalue and eigenvector analysis of the matrix
\begin{equation}\label{cca-mtx}
\bM =\bSigma_\eta^{-1/2}\bSigma_{\eta \eta_m}\bSigma_{\eta_{m}}^{-1}\bSigma_{\eta_m\eta}\bSigma_\eta^{-1/2}.
\end{equation}
It is easy to see that rank($\bM$) = $r$. Furthermore,  let 
$\bxi_{t,m}=(\bxi_{t-1}^\T,\ldots,\bxi_{t-m}^\T)^\T$, 
where again $\bxi_t = (\bff_t^{\T},\bve_t^{\T})^{\T}$, and define $\bSigma_\xi$, 
$\bSigma_{\xi \xi_m}$ and $\bSigma_{\xi_m}$ as the covariance matrices of the given random 
vectors. It is easy to verify that
\[ \bM = \bL\bSigma_{\xi {\xi_m}}\bSigma_{\xi_m}^{-1}\bSigma_{{\xi_m}\xi}\bL^\T=\bL_1\bSigma_{f{\xi_m}}\bSigma_{\xi_m}^{-1}\bSigma_{{\xi_m}f}\bL_1^\T.\]
Let $\lambda_1^2 \geq \lambda_2^2 \geq \cdots \geq \lambda_p^2$ be the 
ordered eigenvalues of $\bM$ and let $[\ba_1,\ba_2,\ldots,\ba_p]$ be the corresponding 
eigenvectors. Then, $\lambda_r \neq 0$, but $\lambda_j = 0$ for $j > r$, and we may take 
$\bL = [\ba_1,\cdots,\ba_p]$, which is an orthonormal matrix. Making use of the properties of 
canonical correlation analysis, we have
\[ \bL^{\T}\bSigma_\eta^{-1/2}\bfeta_t = \bxi_t, \]
and Cov($\bxi_t$) = $\bI_p$. Thus, $\bxi_t = \bL^\T\bSigma_\eta^{-1/2}\bfeta_t$,  
$\bL_1 = [\ba_1,\ldots,\ba_r]$ is the loading matrix of the common factors $\bff_t$, 
and $\bL_2 = [\ba_{r+1},\ldots,\ba_p]$. 
This shows that 
the model in (\ref{cca-transform}) always exists for a VARMA process $\bfeta_t$. 
Furthermore, if the $r$ non-zero eigenvalues of $\bM$ are distinct, 
$\bL_1$ is uniquely defined if we ignore the trivial replacement of $\ba_j$ by $-\ba_j$.

Let 
 \begin{equation}
\label{hat:sge}
\hat{\bSigma}_{\eta}=\frac{1}{T}\sum_{t=1}^{T}(\hat{\bfeta}_{t}-\bar{\bfeta})(\hat{\bfeta}_{t}-\bar{\bfeta})^\T,
 \end{equation}
where $\bar{\bfeta}=\sum_{t=1}^T\hat{\bfeta}_t/T$. The sample estimators $\hat{\bSigma}_{\eta_m}$, $\hat{\bSigma}_{\eta \eta_m}$ are defined in a similar way and the index exceeds $1$ or $T$ are set to be $0$. This leads to the following natural estimator for $\bM$,
\begin{equation}\label{est:m}
\hat{\bM}=\wh\bSigma_\eta^{-1/2}\wh\bSigma_{\eta \eta_m}\wh\bSigma_{\eta_m}^{-1}\wh\bSigma_{\eta_m\eta}\wh\bSigma_\eta^{-1/2}.
\end{equation}
The above discussion also gives rise to an estimator of $\bL$ as 
$\hat{\bL}=(\hat{\ba}_1,...,\hat{\ba}_r,\hat\ba_{r+1},...,\hat\ba_{p})$, where $\hat{\ba}_1,...,\hat{\ba}_r$ are the orthonormal eigenvectors of $\hat{\bM}$ corresponding to the $r$ largest eigenvalues $\hat{\lambda}_1^2\geq\cdots\geq\hat{\lambda}_r^2$, and $\hat{\ba}_{r+1},\ldots,\hat{\ba}_p$ are the orthonormal eigenvectors of $\hat{\bM}$ corresponding to the $p-r$ smallest eigenvalues $\hat{\lambda}_{r+1}^2\geq\cdots\geq\hat{\lambda}_p^2$. Since $\hat{\bL}$ is an 
orthonormal matrix, i.e. $\hat{\bL}^{\T}\hat{\bL} = \bI_p$, 
we may extract the factor process by $\hat{\bff}_t=\hat{\bL}_1^\T\wh\bSigma_\eta^{-1/2}\hat{\bfeta}_t$.

\subsection{Selections of $d_0$, $k_0$, and $r$}
The estimation of $\btheta_i$ in (\ref{lse:th}) assumes $d_0$ and $k_0$ are known, but  these integers must be estimated in practice. We propose to determine $k_0$ and $d_0$ based on the following marginal Bayesian information criterion,
\begin{equation}\label{BIC}
\text{BIC}_i(k,d)=\log [\text{RSS}_i(k,d)/T]+\frac{d+k}{T}C_T\log(p\vee T),\quad i=1, ..., p,
\end{equation}
where $\text{RSS}_i(k,d)$ is similarly defined as (\ref{RSS}) for some $k$ and $d$, $p\vee T=\max(p,T)$, and $C_T>0$ is some constant which diverges together with $T$; see Assumption 3 in Section 3. We often take $C_T$ to be $\log\{\log(T)\}$. Let $\bar{k}$ and $\bar{d}$ be two 
pre-specified  integers and
\begin{equation}\label{kd}
(\hat{k}_i,\hat{d}_i)=\arg\min_{1\leq k\leq\bar{k},1\leq d\leq \bar{d}}\text{BIC}_i(k,d),\quad 
i=1, ..., p.
\end{equation} 
We take $\hat{k}=\max_{1\leq i\leq p}\hat{k}_i$ and $\hat{d}=\max_{1\leq i \leq p}\hat{d}_i$ as the estimators of $k_0$ and $d_0$, respectively. Theorem 2 in Section 3 shows that under some assumptions, $P(\hat{k}=k_0,\hat{d}=d_0)\rightarrow 1$ as $T$ and $p\rightarrow\infty$.

\begin{remark}
In practice, $d_0=1$ or $2$ is often sufficient in characterizing the trend of many time series 
and space-time data. Therefore, we may fix $d_0$ and use the following estimator for $k_0$,
\begin{equation}\label{BIC:ap}
\hat{k}=\max_{1\leq i\leq p}\{\arg\min_{1\leq k\leq \bar{k}}\text{BIC}_i(k,d_0)\}.
\end{equation}
Our numerical study shows that the procedure is insensitive
to the choice of $\bar{k}$ provided $\bar{k}\geq k_0$ and $\bar{k} \leq s/2-1$, which is the maximum possible value to avoid any singularity, or choose $\bar{k}$ by
checking the curvature of $\text{BIC}_i(k)$ directly.
\end{remark}

Turn to the estimation of $r$, which plays an important role in 
the proposed statistical inference. In practice, we may estimate the number of zero canonical correlations $v$ (and hence $r$) by testing the null hypothesis $H_0: \lambda_{p-v+1}^2=\cdots=\lambda_p^2=0$ and $\lambda_{p-v}^2\neq 0$ versus 
the alternative hypothesis $H_a: \lambda_{p-v}^2 = 0$, where $\lambda_i^2$ are the ordered 
eigenvalues of $\bM$.  A test statistic available for testing the hypothesis is 
\begin{equation}\label{test}
S_T(v)=-(T-m+1)\sum_{i=1}^v\log(1-\wh\lambda_{p-i+1}^2). 
\end{equation}
See \cite{TiaoTsay_1989}. 
Under the null hypothesis and some regularity conditions, $S_T(v)$ is distributed as $\chi^2_{v[(m-1)p+v]}$.  Since we allow for the dimension $p$ to increase with the sample 
$T$, we modify the test statistic accordingly by making use of the central limit theorem 
and properties of $\chi^2$ random variables. Specifically, we employ a standardized version 
of the test statistic:
\begin{equation}\label{modified-test}
C_T(v) = \frac{S_T(v)-v[(m-1)p+v]}{\sqrt{2v[(m-1)p+v]}}. 
\end{equation}
Then, under the null hypothesis and some regularity conditions, 
$C_T(v)$ converges in distribution to N(0,1) as $p \rightarrow \infty$. 
Note that the test statistic in Equation (\ref{test}) is for testing the number of 
scalar components of order (0,0) so that there is no need to consider the 
normalization of eigenvalues. Details are given in \cite{TiaoTsay_1989}. 

Using the test statistics in (\ref{test}) or (\ref{modified-test}), one can perform 
the hypothesis testing sequentially 
starting with $v=1$ and until the null hypothesis is rejected. The resulting number of 
zero eigenvalues $\hat{v}$ is an estimate of $v$, and we have $\hat{r} = p-\hat{v}$.

\section{Theoretical Properties}
We present some asymptotic theory for the estimation methods described in Section 2 when $T$, $p\rightarrow\infty$.
We assume $\{(\by_t,\bff_t)\}$ is $\alpha$-mixing with the mixing coefficients defined by
\begin{equation}\label{amix}
\alpha_p(k)=\sup_{i}\sup_{A\in\mathcal{F}_{-\infty}^i,B\in \mathcal{F}_{i+k}^\infty}|P(A\cap B)-P(A)P(B)|,
\end{equation}
where $\mathcal{F}_i^j$ is the $\sigma$-field generated by $\{(\by_t,\bff_t):i\leq t\leq j\}$.
\begin{assumption}\label{a1}
The process $\{(\by_t,\bff_t)\}$ is $\alpha$-mixing with the mixing coefficients satisfying the condition $\sum_{k=1}^\infty\alpha_p(k)^{1-2/\gamma}<\infty$ for some $\gamma>2$, where $\alpha_p(k)$ is defined in (\ref{amix}).
\end{assumption}
\begin{assumption}\label{a2}
For any $i=1,...,p$, $E|\eta_{it}|^{2\gamma}\leq C_1$, where $C_1>0$ is a constant, $\gamma$ is given in Assumption \ref{a1}.
\end{assumption}
Assumption 1 is standard for dependent random sequences. See \cite{gaoetal2017} for a theoretical justification for VAR models. The condition $E|\eta_{it}|^{2\gamma}\leq C_1$ in Assumption 2 can be guaranteed by some suitable conditions on   $\xi_{it}$ and each row of $\wt\bL$ defined in 
Equation (\ref{cca-transform}). 
For example, it holds if $\max_{i,t}E|\xi_{it}|^{2\gamma}<\infty$ and $\max_i\sum_{j=1}^p|\wt\bL_{ij}|<\infty$, where $\wt\bL_{ij}$ is the $(i,j)$-th element of $\wt\bL$. The following theorem establishes the convergence rates of the coefficient estimates for the trend and seasonal parts component-wisely.
\begin{theorem}
If Assumptions 1-2 hold and $k_0$ and $d_0$ are known, as $T,p\rightarrow\infty$, we have
$$|\hat{\alpha}_{ij}-\alpha_{ij}|=O_p(T^{-(2j+1)/2}),\,\,|\hat{\beta}_{il}-\beta_{il}|=O_p(T^{-1/2})\,\,\text{and}\,\,|\hat{\gamma}_{il}-\gamma_{il}|=O_p(T^{-1/2}),$$
for $1\leq i\leq p$, $0\leq j\leq d_0$ and $1\leq l\leq k_0$.
\end{theorem}

Theorem 1 implies that the convergence rates do not depend on the dimension $p$, which is reasonable since the dimension of each $\btheta_i$ is finite. Thus, they are as optimal as the regression estimators with the dimension fixed. To show the consistency of the selected $\wh k$ and $\wh d$ by BIC, we need to impose a condition on the magnitude of the coefficients of the largest orders of the  time trend and seasonal components. 

\begin{assumption}
For each $i=1 ,..., p$, $\alpha_{id_0}^2>>C_T (d_0+k_0)/T\log (p\vee T)$ and $\beta_{ik_0}^2+\gamma_{ik_0}^2>>C_T (d_0+k_0)/T\log (p\vee T)$, where $C_T\rightarrow\infty$ as $T\rightarrow \infty$.
\end{assumption}

Assumption 3 ensures that the orders of the polynomial trend ($d_0$) and the number of the trigonometric series ($k_0$) are asymptotically identifiable in the sense that their 
coefficients can be detected as non-zero ones  as $\{T^{-1}\log(p\vee T)\}^{1/2}$
is the minimum order of a non-zero coefficient to be identifiable;
see, for instance, Luo and Chen (2013). 

We now state the consistency of the selectors $\wh k$ and $\wh d$ defined in (\ref{kd}).
\begin{theorem}
If Assumptions 1-3 hold, then $P(\hat{k}=k_0,\hat{d}=d_0)\rightarrow 1$ as $T, p\rightarrow\infty$.
\end{theorem}

Theorem 2 implies that we can consistently estimate $d_0$ and $k_0$ under some regularity conditions as the dimension $p$ and the sample size $T$ go to infinity. Therefore, we can replace $d_0$ and $k_0$ by $\wh d$ and $\wh k$, respectively, in the estimators $\btheta_i$ in Section 2.
To establish the results for estimating factor loadings, we introduce more assumptions.

\begin{assumption}
There exist positive constants $C_2$, $C_3$, $C_4$, $\kappa_1$ and $\kappa_2$ such that $\|\bSigma_{f{\xi_m}}\|_{2}\leq  C_2 $ and $C_3 \leq \|\bSigma_{\xi_m}\|_{\min}\leq \|\bSigma_{\xi_m}\|_{2}\leq C_4$, and $\kappa_1\leq \lambda_{\min}(\wt\bL\wt\bL^\T)\leq\lambda_{\max}(\wt\bL\wt\bL^{\T})\leq \kappa_2$, where $\kappa_1$ and $\kappa_2$ may diverge in relation to the dimension $p$.
\end{assumption}

\begin{assumption}
The matrix $\bM$ admits $r$ distinct positive eigenvalues $\lambda_1^2>\cdots>\lambda_r^2>0$.
\end{assumption}

The boundedness  condition of the  eigenvalues of $\bSigma_{f\xi_m}$ and $\bSigma_{\xi_m}$ in Assumption 4 is natural since we have standardized $\bxi_t$ to $\Cov(\bxi_t)=\bI_p$,
and  is related to the dimension $p$ since the columns of $\bL_1$ are standardized. 
$\kappa_1$ and $\kappa_2$ of Assumption 4 control the strength of the transformation matrix $\wt\bL$.
Assumption 5 implies that $\bL_1$ defined in Section 2.2 is unique.
This simplifies the presentation significantly,
as Theorem~\ref{tm1} below can present the convergence rates of
the estimator for $\bL_1$ directly. Without Assumption 5, the same convergence
rates can be obtained for the estimation of the linear space
$\mathcal{M}(\widehat{\bL}_1)$; see (\ref{ratesSP}) in Theorem 4 below.

\begin{theorem}
\label{tm1} If Assumptions 1-5 hold and suppose that
$r$ is known and fixed, then
\[
\|\widehat{{\bf L}}_i-{\bf L}_i\|_2 =~\left\{
     \begin{array}{ll}
       O_p(
T^{-1/2}), & {\rm if\; \;}  p \text{\rm \;\; is fixed};
 \\[1ex]
       O_p(\kappa_1^{-2}\kappa_2pT^{-1/2}), \quad &{\rm if\;} p=o\{\min(T^{1/2},\kappa_2^{-1}\kappa_1^2T^{1/2})\}; i=1,2.
     \end{array}
   \right.
\]
Furthermore, 
\begin{align*}
 &p^{-1/2}\|\widehat{{\bf L}}_1\widehat{{\bf f}}_t-{\bf L}_1{\bf
f}_t\|_2\\
&=\left\{
     \begin{array}{ll}
       O_p(T^{-1/2}),&{\rm if}\;
p\; \text{\rm is fixed};
 \\[1ex]
       O_p(\|\wh\bL_1-\bL_1\|_2+\kappa_1^{-3/2}p^{3/2}T^{-1/2}+\kappa_1^{-1/2}p^{1/2}T^{-1/2}), \quad&{ \rm if} \;
p=o\{\min(T^{1/2},\kappa_2^{-1}\kappa_1^2T^{1/2})\}.
     \end{array}
   \right.
\end{align*}
\end{theorem}

From Theorem 3 and, as expected, the convergence rates  are all standard at $\sqrt{T}$, which is commonly seen in the traditional statistical theory. When $p$ is diverging, the upper bounds in Theorem 3 are all $pT^{-1/2}$ if we assume $\kappa_1$ and $\kappa_2$ are finite, implying  
that the condition $p=o(T^{1/2})$ is needed to guarantee the consistency. On the other hand, if $\kappa_1\asymp\kappa_2\asymp T^\iota$ for some $0<\iota<1$,  we have $\|\widehat{{\bf L}}_i-{\bf L}_i\|_2=O_p(p^{1-\iota}T^{-1/2})$ and $p^{-1/2}\|\widehat{{\bf L}}_1\widehat{{\bf f}}_t-{\bf L}_1{\bf
f}_t\|_2=O_p(p^{3(1-\iota)/2}T^{-1/2})$. Therefore, we need $p=o(T^{1/(3(1-\iota))})$ in this case. For example, if $\iota=1/4$, then the condition becomes $p=o(T^{4/9})$ to guarantee the consistency.


In general, the choice of $\bL_1$ in Model (\ref{factor}) is not
unique so we consider the error in estimating $\mathcal{M}(\bL_1)$ instead of
a particular $\bL_1$, because $\mathcal{M}(\bL_1)$ is uniquely defined by
(\ref{factor}) and it does not vary with different choices of
$\bL_1$. To this end, we adopt the discrepancy measure used by
\cite{panyao2008}: for two $p\times r$ half orthogonal
matrices ${\bf H}_1$ and ${\bf H}_2$ satisfying the condition ${\bf
H}_1^\T{\bf H}_1={\bf H}_2^\T{\bf H}_2=\bI_{r}$, the difference
between the two linear spaces $\mathcal{M}({\bf H}_1)$ and
$\mathcal{M}({\bf H}_2)$ is measured by
\begin{equation}
D(\mathcal{M}({\bf H}_1),\mathcal{M}({\bf
H}_2))=\sqrt{1-\frac{1}{r}\textrm{tr}({\bf H}_1{\bf H}_1^\T{\bf
H}_2{\bf H}_2^\T)}.\label{eq:D}
\end{equation}
Note that $D(\mathcal{M}({\bf H}_1),\mathcal{M}({\bf H}_2))$ always
assumes values between 0 and 1. It is equal to $0$ if and only if
$\mathcal{M}({\bf H}_1)=\mathcal{M}({\bf H}_2)$, and to $1$ if and
only if $\mathcal{M}({\bf H}_1)\perp \mathcal{M}({\bf H}_2)$.  The following theorem establishes the convergence of $D(\mathcal{M}(\widehat{\bf L}_1),\mathcal{M}({\bf L}_1))$ when $\bL_1$ is not uniquely defined.

\begin{theorem}
\label{tm3} Suppose Assumptions 1-4 hold. Assume that
$r$ is known and fixed, then for $i=1, 2$,
\begin{equation} \label{ratesSP}
D(\mathcal{M}(\widehat{\bf L}_i),\mathcal{M}({\bf L}_i)) =~\left\{
     \begin{array}{ll}
       O_p(
T^{-1/2}), & {\rm if\; \;}  p \text{\rm \;\; is fixed}; \\[1ex]
       O_p(\kappa_1^{-2}\kappa_2pT^{-1/2}), \quad &{\rm if\;} p=o\{\min(T^{1/2},\kappa_2^{-1}\kappa_1^2T^{1/2})\}.
     \end{array}
   \right.
\end{equation}
\end{theorem}

The results of Theorems 3 and 4 are obtained when the number of factors $r$ is given. 
In practice, $r$ requires estimation. To this end, we first state the asymptotic properties of the test statistic defined in (\ref{test}).

\begin{theorem}
\label{tm4} Suppose Assumptions 1-4 hold.

(i) If $p$ is fixed, under $H_0$, the statistic $S_T(v)$ converges to a chi-squared random variable with degrees of freedom $v[(m-1)p+v]$ as $T\rightarrow\infty$.

(ii) If $p=o\{\min(T^{1/2},\kappa_2^{-1}\kappa_1^2T^{1/2})\}$, under $H_0$, 
$$\frac{S_T(v)-v[(m-1)p+v]}{\sqrt{2v[(m-1)p+v]}}\rightarrow_d N(0,1),$$
as $T\rightarrow\infty$.
\end{theorem}

The next theorem establishes the consistency of the estimator $\wh v$ defined in Section 2.
\begin{theorem}
Let Assumptions 1-4 hold. Under the null hypothesis that $l=v$,

(i) if $p$ is fixed, $S_T(l)$ diverges to infinity as $T\rightarrow\infty$ for any $l>v$.

(ii) if $p=o\{\min(T^{1/2},\kappa_2^{-1}\kappa_1^2T^{1/2})\}$, $\frac{S_T(l)-l[(m-1)p+l]}{\sqrt{2l[(m-1)p+l]}}$ diverges to infinity as $T\rightarrow\infty$ for any $l>v$.

Therefore, our test of the null hypothesis  of $l = v$ versus
the alternative of $v <l\leq p$ is consistent and has the asymptotically correct size.
\end{theorem}

Theorems 5 and 6 together imply that we can consistently estimate the number of factors $r$. With the estimator $\wh r (=p-\wh v)$, we may define an estimator for $\bL_1$ as
$\wh \bL_1 = (\wh \ba_1, \ldots, \wh \ba_{\wh r} )$, where $\wh \ba_1,
\ldots, \wh \ba_{\wh r}$ are the orthonormal eigenvectors of $\wh
\bM$, defined in (\ref{est:m}), corresponding to the $\wh r$
largest eigenvalues.  To
measure the error in estimating the factor loading space, we use
\begin{equation}\label{dm}
\widetilde{D}(\mathcal{M}(\wh{{\bf L}}_1),\mathcal{M}({\bf
L}_1))=\sqrt{1-\frac{1}{\min(\wh r,r)}\textrm{tr}(\wh{{\bf
L}}_1\wh{{\bf L}}_1^\T{\bf L}_1{\bf L}_1^\T)}.
\end{equation}
which is a modified version of (\ref{eq:D}), and it allows the dimensions of $\mathcal{M}(\wh \bL_1)$ and $\mathcal{M}(\bL_1)$ to be different.

\begin{remark}
(i) Our method and theory can be extended to the cases when $\bff_t$ in Model (\ref{cca-transform})  is unit-root non-stationary and hence $\bfeta_t$ is non-stationary. A simple condition is that a generalized sample (auto)covariance matrix
\[T^{-\delta}\sum_{t=1}^{T-k}(\bff_{t+k}-\bar{\bff})(\bff_t-\bar{\bff})^\T\]
converges weakly, where $\delta>1$ is a constant. This weak convergence has been established when, for example, $\{\bff_t\}$ is an integrated process of order $2$ by \cite{pena2006}. \\
(ii) On the other hand, if $\bff_t$ and, hence, $\bfeta_t$ is non-stationary, we can replace the definition of $\bSigma_\eta (k)=\Cov(\bfeta_{t+k},\bfeta_t)$ by
\[\bSigma_\eta (k)=\frac{1}{T-k}\sum_{t=1}^{T-k}\Cov(\bfeta_{t+k},\bfeta_t),\]
and similarly for others. All the theory still works under the mixing condition in Assumption 1;  
see the argument in \cite{changguoyao2015} for details, but we do not pursue the issue further.
\end{remark}

\section{Numerical Properties}
\subsection{Simulation Studies}
In this section, we illustrate the finite-sample performance of the proposed method via 
simulation. 
The data generating process is 
\begin{equation}\label{ex1}
\by_t=\bTheta\bd_t+\bfeta_t\quad \text{and}\quad \bfeta_t=\wt\bL\left[\begin{array}{c}
\bff_t\\ \bve_t
\end{array}\right],
\end{equation}
where $\bTheta=(\btheta_1,...,\btheta_p)^\T$, and $\btheta_i$ and $\bd_t$ are defined in (\ref{lse}). We set the periodicity $s=30$, the true number of the trigonometric series 
$k_0=5, 8, 10$, the number of factors $r=3$, the dimension $p=10, 15, 30, 50$, and the sample size $T=200, 500, 1000, 2000, 3000$, respectively. $\bve_t\sim N(0,\bI_{p-r})$, and $\bff_t$ follows the \text{VAR}(1) model:
\begin{equation}
\bff_t=\bPhi\bff_{t-1}+\bu_t,
\end{equation}
where $\bPhi$ is a diagonal matrix with its diagonal elements drawn randomly 
from $U(0.2,0.9)$, $\bu_t\sim N(0,\bI_r)$, 
and the elements of $\bTheta$ and $\wt\bL$ are drawn independently from $U(-2,2)$ for each setting and replication.  In view of Remark 1, 
we always treat $d_0$ as known and consider $d_0=1$ or $2$ since a larger one is not of 
interest in practice. $\bar{k}$ is set to be $14$ $(=s/2-1)$ in (\ref{BIC:ap}).  We use $1000$ replications in each experiment.

For the performance of the estimator in (\ref{BIC:ap}), we set $d_0=1$, because the choice of $d_0=2$ gives similar results. The empirical probabilities $P(\wh k=k_0)$ are reported in 
Table~\ref{Table1}. From 
the table, we see that, for a given $(p,k_0)$, performance of the proposed method improves as the sample size increases. On the other hand, for a given $(k_0,T)$, the empirical probability 
of correct selection decreases slightly as $p$ increases.  This is reasonable since it is harder to locate the number of basis functions when the dimension becomes higher. Overall, the 
proposed method works well even in the case of a small sample size (e.g. $T=200$) and  high dimension (e.g. $p=50$).

\begin{table}
\begin{center}
 \setlength{\abovecaptionskip}{0pt}
\setlength{\belowcaptionskip}{3pt}
 \caption{Empirical probabilities $P(\hat{k}=k_0)$ of Model (\ref{ex1}) with $d_0=1$, where $p$ and $T$ are the dimension and sample size, respectively, and 
 $k_0$ is the number of trigonometric components. 1000 iterations are used.} \centering
          \label{Table1}
\begin{tabular}{cc|ccccc}\hline 
& & \multicolumn{5}{|c}{$T$} \\ \hline
$p$ & $k_0$ &$200$&$500$&$1000$&$2000$&$3000$\\
\hline
&5&0.912&0.986&0.998&1&1\\
10&8&0.956&0.994&0.996&1&1\\
&10&0.960&0.994&0.998&1&1\\
\hline
&5&0.908&0.976&0.994&0.998&1\\
15&8&0.920&0.988&0.994&1&1\\
&10&0.932&0.988&0.996&0.998&1\\
\hline
&5&0.888&0.936&0.982&0.990&0.998\\
30&8&0.886&0.956&0.974&0.992&0.996\\
&10&0.858&0.958&0.980&1&1\\
\hline
&5&0.848&0.932&0.974&0.986&0.996\\
50&8&0.640&0.952&0.966&0.978&0.994\\
&10&0.668&0.960&0.974&0.982&0.998\\
   \hline 
\end{tabular}
          \end{center}
\end{table}

Next, we study the performance of the test statistic defined in (\ref{test}). We take $m=2$ and only show the results for $d_0=1$ and $k_0=5$, because similar results are obtained for different configurations of $m$, $d_0$ and $k_0$. Furthermore, we also compare the 
proposed estimator with the ratio-based estimator
\begin{equation}\label{rto}
\hat{r}=\arg\min_{1\leq j\leq p-1}\frac{\hat{\lambda}_{j+1}}{\hat{\lambda}_j},
\end{equation}
which was used in Lam and Yao (2012) to determine the number of common factors. 
Table~\ref{Table2} provides the  empirical probabilities $P(\wh r=r)$ of Model (\ref{ex1}) for $r=3$. 
The results for  the cases of $r=5$ and $8$ are displayed in Tables 1-2 of the online supplement,  
because the results are similar. For almost every setting of  $(p,T)$, our method based on the test statistic in (\ref{test})  fares better than the one based on the ratio in (\ref{rto}).  It also shows that for both methods the impacts in estimating $r$ caused by errors in estimating $k_0$ is almost negligible.  In addition, the performance of our method improves as the sample size increases 
in each setting. On the other hand, 
the performance of both methods deteriorates for a fixed sample size when 
the dimension $p$ increases. This is understandable because our test statistic depends on the consistency of the covariance matrix 
estimator, which requires $p=o(T^{1/2})$, and the estimator in (\ref{rto}) becomes more variable. 
\begin{table}
\begin{center}
 \setlength{\abovecaptionskip}{0pt}
\setlength{\belowcaptionskip}{3pt}
 \caption{The empirical probabilities $P(\hat{r}=r)$ of Model (\ref{ex1}) with $d_0=1$ and $k_0=5$, where $p$ and $T$ are the dimension and sample size, respectively. `Test' and `Ratio' denote the estimators determined by the test statistic in (\ref{test}) and (\ref{rto}), respectively. 
 1000 iterations are used.} \centering
          \label{Table2}
\begin{tabular}{ccccccccc}
\hline 
&&$T$&&$200$&$500$&$1000$&$2000$&$3000$\\
\hline
$r=3$&$k_0$ known&$p=10$&Test&0.447&0.703&0.864&0.935&0.943\\
&&&Ratio&0.088&0.350&0.595&0.781&0.863\\
&&$p=15$&Test&0.333&0.603&0.806&0.908&0.921\\
&&&Ratio&0.023&0.250&0.469&0.672&0.783\\
&&$p=30$&Test&0&0.151&0.527&0.776&0.859\\
&&&Ratio&0&0.096&0.260&0.466&0.549\\
&&$p=50$&Test&0&0&0.024&0.408&0.648\\
&&&Ratio&0&0.018&0.164&0.313&0.396\\
&$k_0$ unknown&$p=10$&Test&0.476&0.722&0.872&0.938&0.941\\
&&&Ratio&0.093&0.365&0.579&0.738&0.852\\
&&$p=15$&Test&0.351&0.611&0.795&0.899&0.905\\
&&&Ratio&0.018&0.242&0.465&0.668&0.778\\
&&$p=30$&Test&0&0.140&0.512&0.763&0.844\\
&&&Ratio&0&0.098&0.269&0.463&0.551\\
&&$p=50$&Test&0&0&0.032&0.352&0.612\\
&&&Ratio&0&0.026&0.154&0.310&0.418\\
\hline
\end{tabular}
          \end{center}
\end{table}

\begin{figure}
\begin{center}
\subfigure[]{\includegraphics[width=0.517\textwidth]{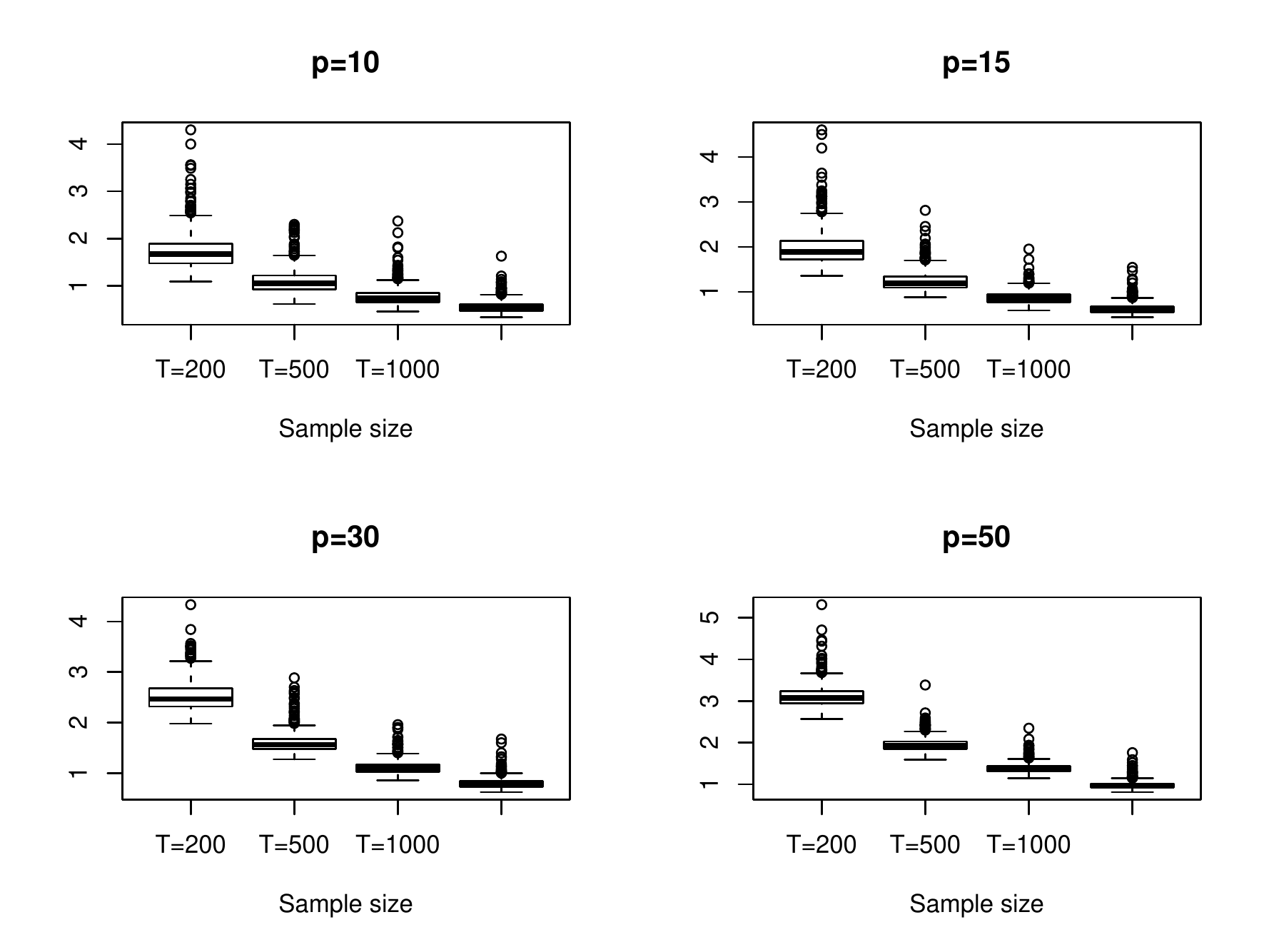}}
\subfigure[]{\includegraphics[width=0.47\textwidth]{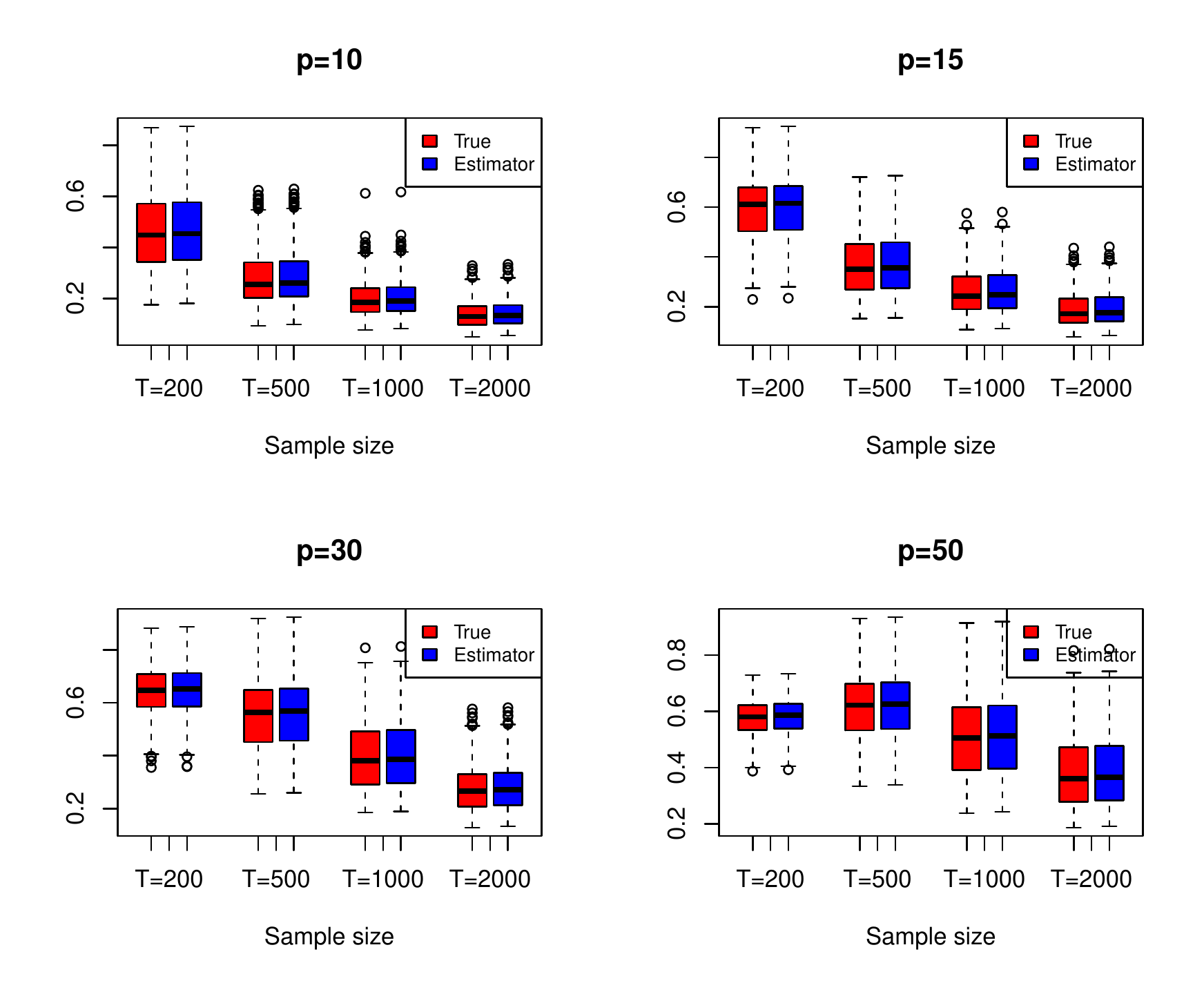}}
\caption{(a) Boxplots of $p^{-1/2}\|\wh\bTheta-\bTheta\|_F$ of Model (\ref{ex1}) with $k_0=5$, $r=3$ and $d_0=1$; (b) Boxplots of $\bar{D}(\mathcal{M}(\wh\bL_1),\mathcal{M}(\bL_1))$ in (\ref{ex1}) with $r=3$, $k_0=5$ and $d_0=1$, where `True' means we use $k_0$ and `Estimator' corresponds using $\wh k$.}\label{fig1}
\end{center}
\end{figure}

We then study the estimation error of the coefficients $\bTheta$ when $k_0=5$. For the cases of 
correctly identified $k_0$, Figures \ref{fig1}(a) reports the boxplots of $p^{-1/2}\|\wh \bTheta-\bTheta\|_F$ for different $p$ and $T$. From Figure~\ref{fig1}(a), 
we  see that the estimation errors decrease as the sample size increases. 
The estimation errors for other settings are  similar and, hence, are omitted.

Finally, we report the boxplots of $D(\mathcal{M}(\wh\bL_1),\mathcal{M}(\bL_1))$ in Figure \ref{fig1}(b). We only consider the case when $r=3$, $d_0=1$ and $k_0=5$ because the results are similar for other cases. Since $\wt\bL$ is not an orthogonal matrix in the simulation models, we  extend the discrepancy measure
in (\ref{dm}) to a more general form below. Let $\bH_i$ be a
$p\times r_i$ matrix with rank$(\bH_i) = r_i$, and $\bP_i =
\bH_i(\bH_i'\bH_i)^{-1} \bH_i'$, $i=1,2$. Define
\begin{equation}\label{dmeasure}
\bar{D}(\mathcal{M}(\bH_1),\mathcal{M}(\bH_2))=\sqrt{1-
\frac{1}{\min{(r_1,r_2)}}\textrm{tr}(\bP_1\bP_2)}.
\end{equation}
Then $\bar{D} \in [0,1]$. Furthermore,
$\bar{D}(\mathcal{M}(\bH_1),\mathcal{M}(\bH_2))=0$ if and only if
either $\mathcal{M}(\bH_1)\subset \mathcal{M}(\bH_2)$ or
$\mathcal{M}(\bH_2)\subset \mathcal{M}(\bH_1)$, and  it is 1 if and only if
$\mathcal{M}(\bH_1) \perp \mathcal{M}(\bH_2)$.
When $r_1 = r_2=r$ and $\bH_i'\bH_i= \bI_r$,
$\bar{D}(\mathcal{M}(\bH_1),\mathcal{M}(\bH_2))
$ is the same as that in (\ref{eq:D}). In the simulation, we take  $\bL_1=\wh\bSigma_\eta^{-1/2}\wt\bL_1$, where $\wh\bSigma_\eta$ is the sample covariance matrix of $\bfeta_t$, 

From Figure \ref{fig1}(b), for each $p$, the discrepancy decreases as the sample increases. The effect of the estimator $\wh k$ is almost negligible, especially when the sample size is large. 
For $p=50$, the discrepancy is smaller and less variable when $T=200$ than that when $T = 500$. 
 This is understandable because the test statistic used to determine the number of 
 common factors $r$ tends to overestimate the true value when the dimension is high and 
 the sample size is small.   Consequently,  the space $\mathcal{M}(\wh\bL_1)$ may cover 
 a larger space than $\mathcal{M}(\bL_1)$. 
 Overall, the simulation results are in line with the asymptotic results obtained in Section 3. 
 In applications, to mitigate the impact of mis-identifying $r$, one can test the serial correlations 
 of the estimated common factor associated with the $\hat{r}$-th largest eigenvalue and adjust 
 $\hat{r}$ accordingly, because under the proposed model common factors are not 
 white noise. See the two examples in the next section.

\subsection{Applications}
\begin{figure}
\begin{center}
\includegraphics[width=5.8in,height=2.3in]{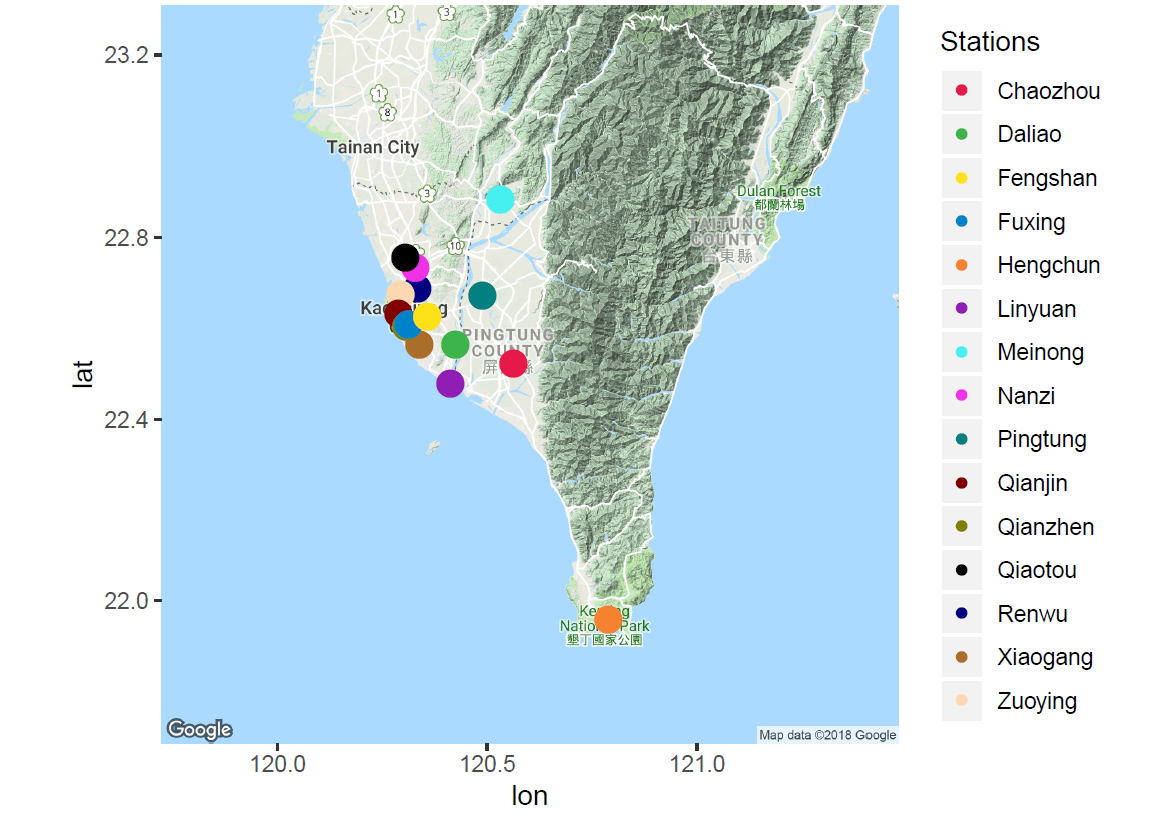}
\end{center}
\caption{Locations (latitude versus longitude) of the 15 monitoring stations in the southern 
part of Taiwan around the industrial city Kaohsiung. The orders of the stations on the right legend are 14, 1, 13, 11, 9, 5, 10, 12, 8, 6, 7, 15, 3, 2, 4, respectively. }\label{location}
\end{figure}

\begin{figure}
\begin{center}
{\includegraphics[width=0.70\textwidth]{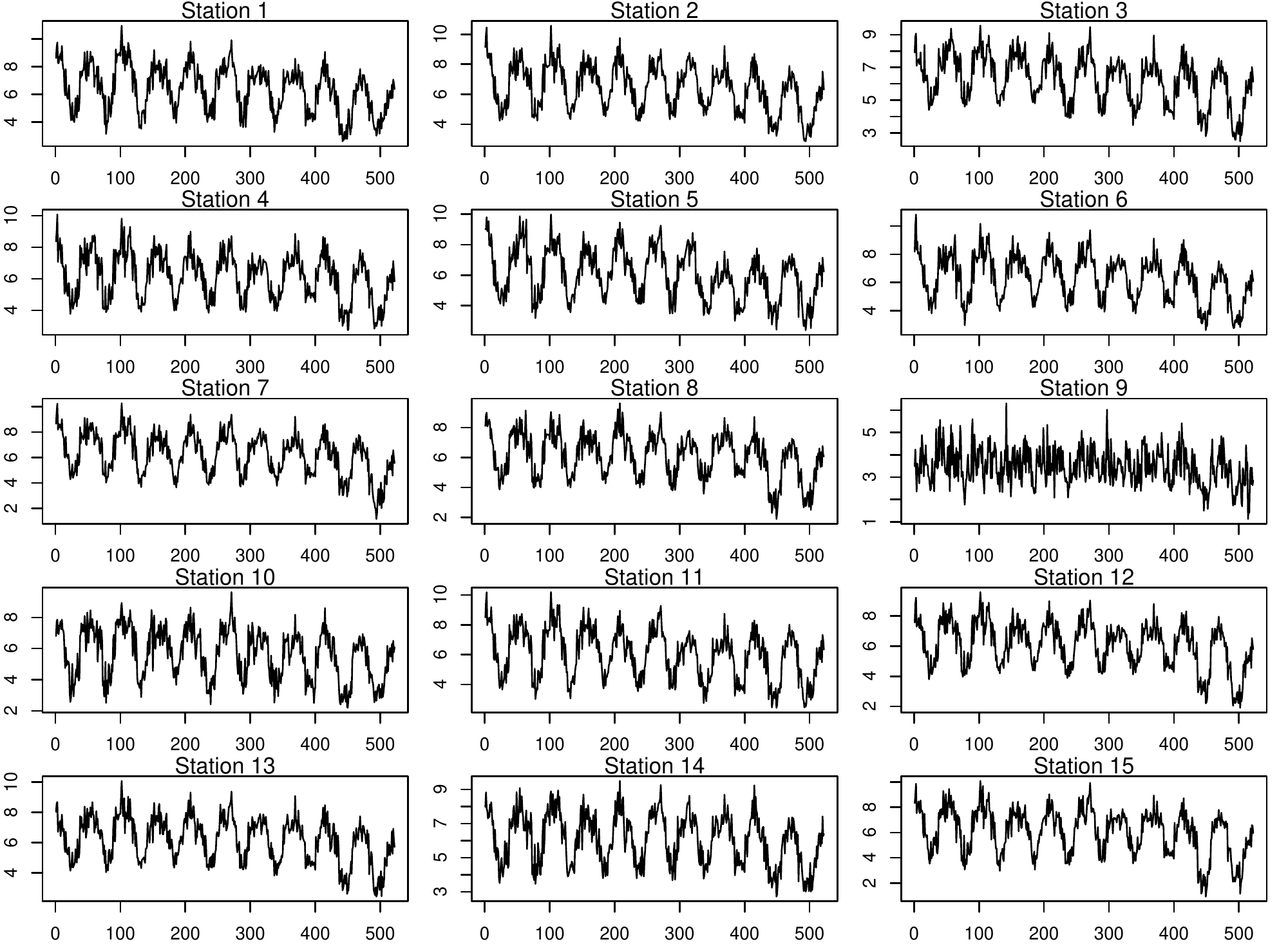}}
\caption{The weekly averages of the square-root-transformed PM$_{2.5}$ of 15 stations 
in southern Taiwan  
from January 2006 to 
December 2015 .}\label{pm25plot}
\end{center}
\end{figure}

{\bf Example 1.}
In this example, we apply the proposed method to modeling a 15-dimensional series of $\text{PM}_{2.5}$ measurements. 
The original PM$_{2.5}$ data were hourly measurements at 15 monitoring stations in the 
southern part of Taiwan 
 from January 1, 2006, to December 31, 2015. 
The locations of the 15 stations are shown in Figure~\ref{location}.
For simplicity, we aggregate the data to weekly observations by taking the average of 
measurements within each week, and then take a square-root transformation. Figure \ref{pm25plot} shows the time plots of the transformed weekly data of  the 15 stations. 
From Figure \ref{pm25plot}, we see clearly that the data possess strong seasonal patterns with periodicity $s\approx 52$. The plots 
also show certain common characteristics across the stations, but some marked differences also 
exist.

\begin{figure}
\begin{center}
{\includegraphics[width=0.70\textwidth]{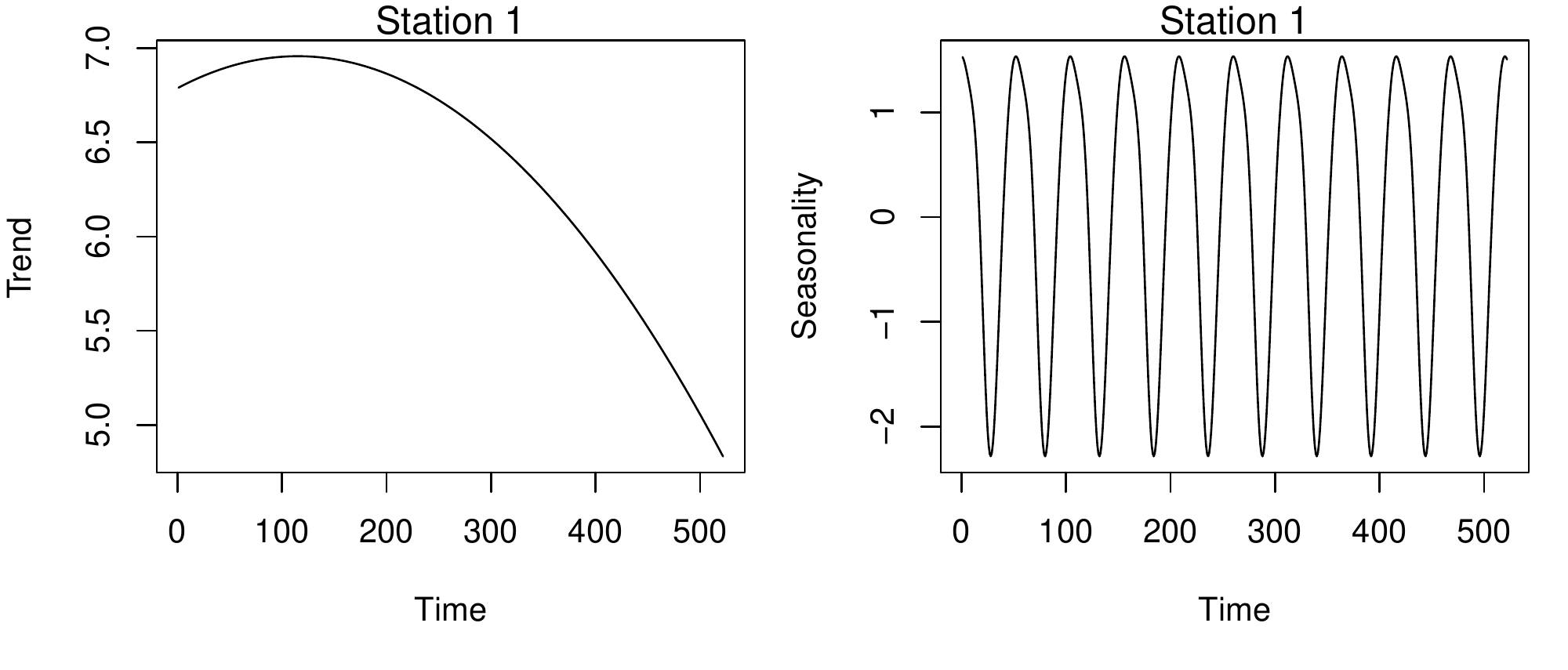}}
\caption{The estimated trend and seasonal components for Station 1 of the square-root-transformed PM$_{2.5}$ in southern Taiwan from January 2006 to December 2015.}\label{fig5}
\end{center}
\end{figure}

To apply the proposed method, we set $d\in\{0,1,2\}$ and $k\in\{1,...,25\}$, and found that $\wh d=2$ and $\wh k=3$ by the method of (\ref{kd}).
Figures \ref{fig5}  plots the estimated trend and seasonal components of Station 1: Daliao, and the others are shown in the supplementary material because they are similar to each other. 
From Figure~\ref{fig5}, we see that  the $\text{PM}_{2.5}$ index was increasing from 
January 2006 to  roughly the 
middle of 2009, and then decreasing after that for all 15 stations. This is because the air pollution had  become a serious issue in recent years and many governments started to issue regulations to 
reduce the emissions after 2005. The Environmental Protection Administration in Taiwan 
also drew up a plan to reduce  PM$_{2.5}$ pollutant starting in 2009, which resulted in a 
slightly  decreasing trend. From Figure~\ref{fig5}, it is clear that the measurements of the $\text{PM}_{2.5}$ are usually high during the winter and 
low in the summer. This is reasonable 
given that the measurement of $\text{PM}_{2.5}$  depends on the consumption of fossil fuels, 
wind direction and speed, and humidity. Winter is the dry season in Taiwan with north-west wind from the Mainland of China. On the contrary, wind is typically from the south-east (Pacific Ocean) during 
the summer with more rains (even typhoons). 
The seasonal patterns of Station 9  (Hengchun) and 10 (Meitong)  are rather different from the other 
stations. See Figure 2 in the online supplement. This is also understandable, 
becuase these two stations are located at rural areas whereas 
the other stations are more close to the industrial city Kaohsiung. 
See Stations `9' and `10' in Figure~\ref{location}.

\begin{figure}
\begin{center}
{\includegraphics[width=0.70\textwidth]{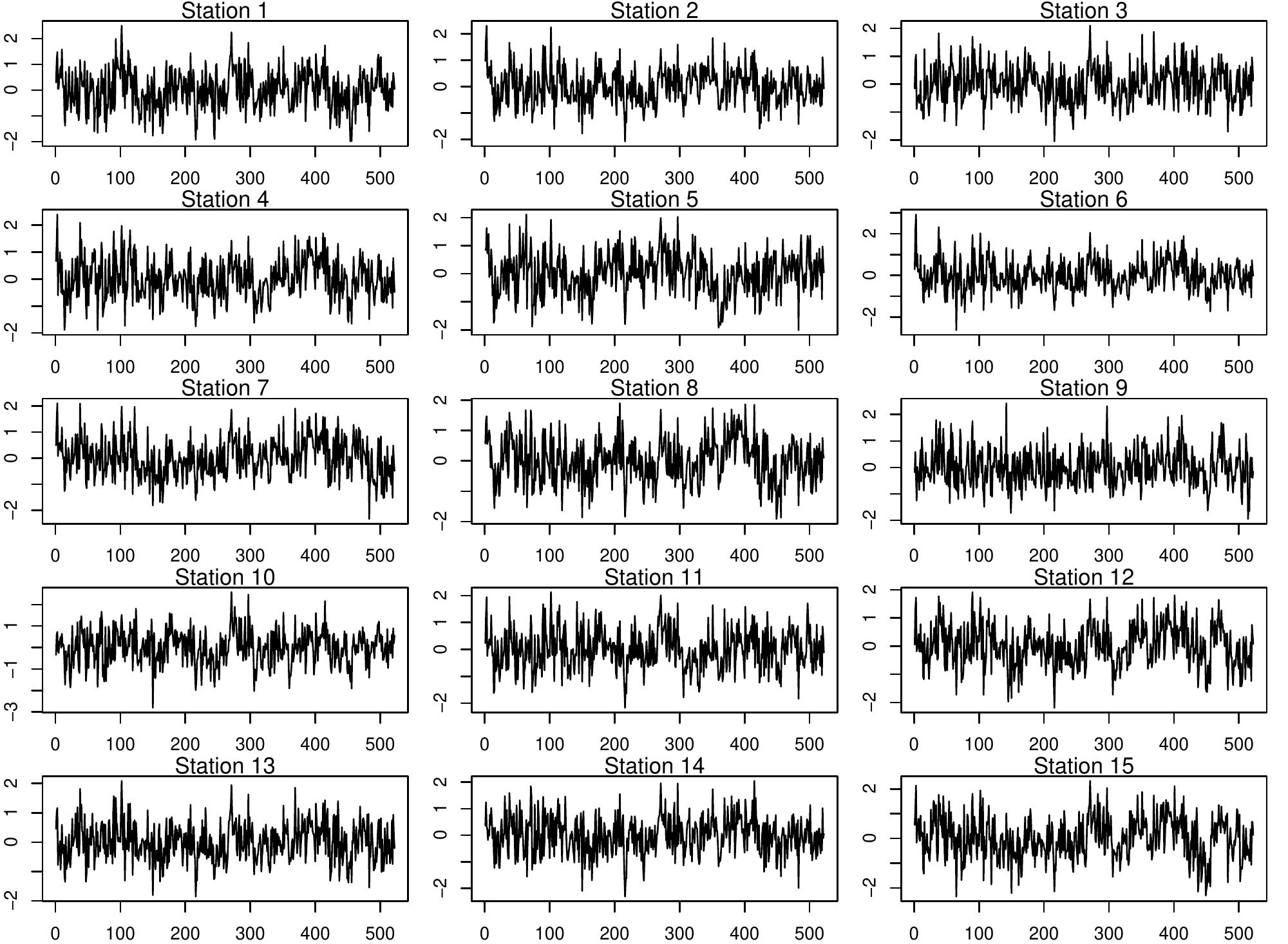}}
\caption{The estimated irregular components for 15 stations of the square-root-transformed PM$_{2.5}$ in southern Taiwan from January 2006 to December 2015.}\label{fig6}
\end{center}
\end{figure}

Figure~\ref{fig6} presents the irregular components after removing the trend and seasonal parts. 
From the plots, we see that the proposed trigonometric series work well 
in modeling the seasonal patterns of the data.  We then apply the proposed method 
to seek common factors in the seasonally adjusted series. For this particular instance, 
we chose $m=2$ for Equation (\ref{est:m}). By applying the proposed test statistic to 
the estimated irregular components, we find that $\wh r=12$,  i.e. we have $12$ estimated latent 
common factors. 
Specifically, the test statistics of (\ref{test}) and their $p$-values, in parentheses, for the first few 
smallest canonical correlations are 12.43(0.71), 36.11(0.37), 72.09(0.051), and 154.17(0.00). 
Therefore, there are three scalar components of order (0,0) among the 15 stations.
The estimated transformation matrix $\wh\bL$ multiplied by 100 is shown in Equation (B.1) of the online supplement.

The time plots the 15 canonical variates and their sample  autocorrelation functions 
are shown in the online supplement, from which we can further confirm that there are 
three white noise series. 
To model the dynamics of the whole 15 stations, we can employ a VARMA model
for the selected 12 latent factors. The resulting vector structural model can then be 
used for prediction. It turns out a VAR(1) model is sufficient for the estimated common 
factor series $\hat{\bff}_t$. Details of the fitted model are omitted.


\begin{figure}
\begin{center}
{\includegraphics[width=0.70\textwidth]{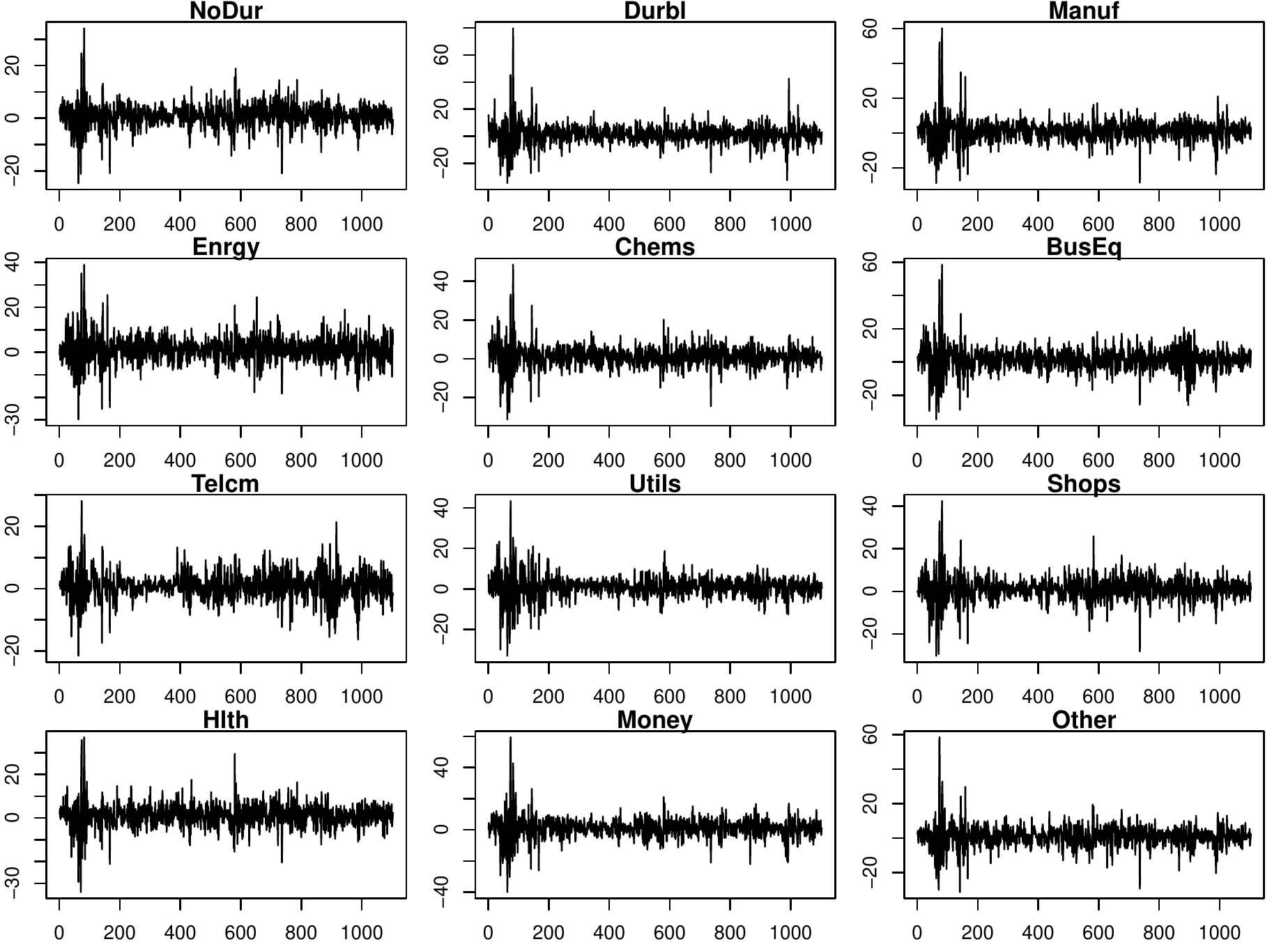}}
\caption{Monthly value-weighted returns of 12 industrial portfolios from July 1926 to May 2018. They are 1. Consumer NonDurables; 2. Consumer Durables; 3. Manufacturing; 4. Energy; 5. Chemicals and Allied Products; 6. Business Equipment; 7. Telephone and Television Transmission; 8. Utilities; 9. Wholesale, Retail, and Some Services; 10. Healthcare, Medical Equipment, and Drugs; 11. Finance; 12. Other.  }\label{fig8}
\end{center}
\end{figure}

{\noindent\bf Example 2.} This example considers the data of monthly value-weighted returns of 12 Industrial Portfolios, which are available at 
\url{http://mba.tuck.dartmouth.edu/pages/faculty/ken.french/data_library.html}. The data  series span from July 1926 to May 2018 with a total of 1103 observations. The 
industrial sectors include the Consumer NonDurables,  Consumer Durables,  Manufacturing,  Energy,  Chemicals and Allied Products,  Business Equipment,  Telephone and Television Transmission,  Utilities, Wholesale, Retail, and Some Services,  Healthcare, Medical Equipment, and Drugs,  Finance, and Others. See Figure \ref{fig8} for the time plots of the data. Therefore, we  have $T=1103$, $p=12$, and the periodicity $s=12$.

\begin{figure}
\begin{center}
{\includegraphics[width=0.70\textwidth]{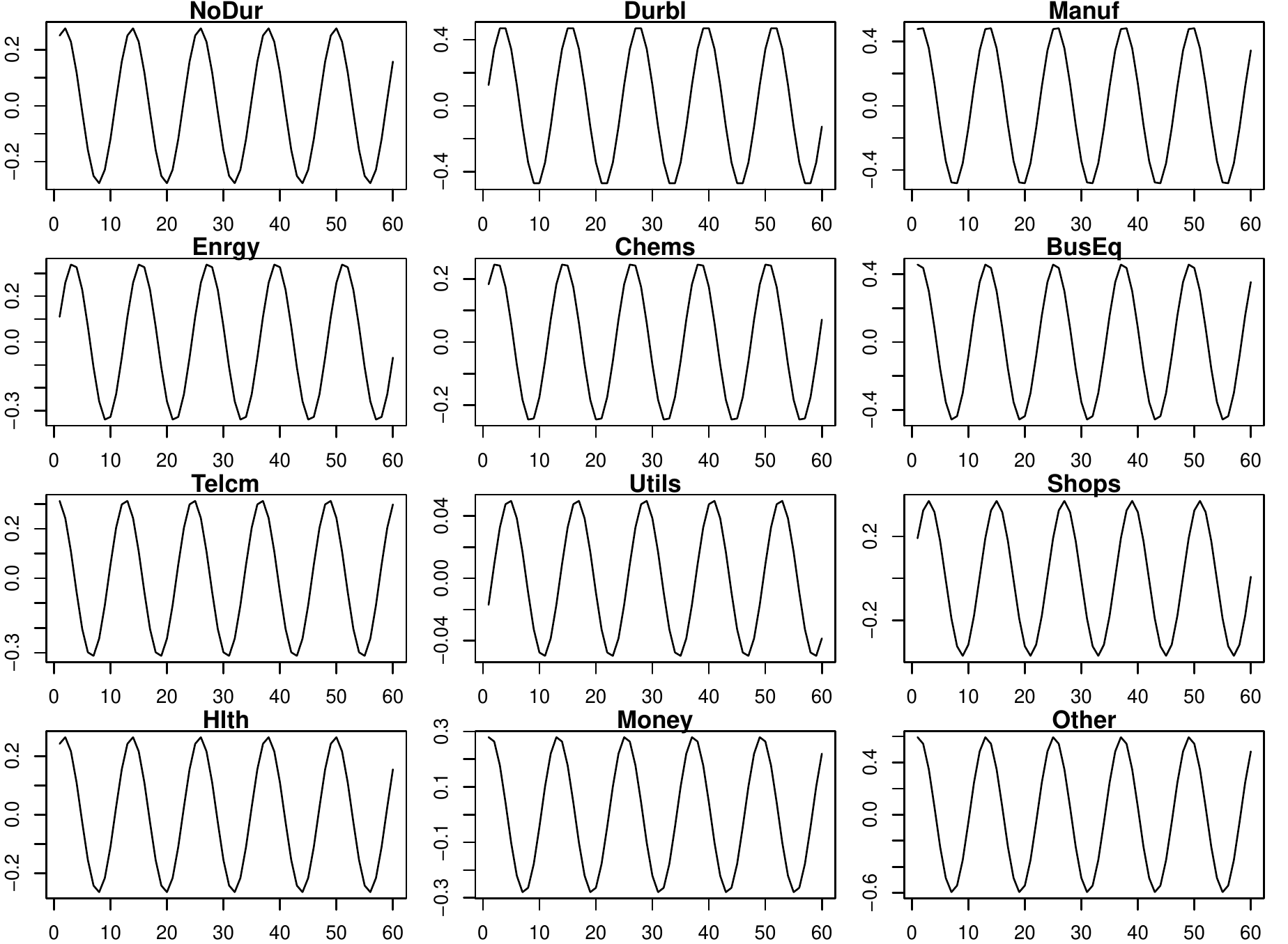}}
\caption{Estimated seasonal components of 12 industrial portfolios from January 2013 to December 2017. They are 1. Consumer NonDurables; 2. Consumer Durables; 3. Manufacturing; 4. Energy; 5. Chemicals and Allied Products; 6. Business Equipment; 7. Telephone and Television Transmission; 8. Utilities; 9. Wholesale, Retail, and Some Services; 10. Healthcare, Medical Equipment, and Drugs; 11. Finance; 12. Other. }\label{fig9}
\end{center}
\end{figure}

We first apply the proposed BIC to the data and found $\wh d=0$ and $\wh k=1$, that is, 
as expected, there is no significant trend in the data. For ease in exposition, we only show the estimated seasonal parts of the 12 industrial portfolios from January 2013 to December 2017 in Figure \ref{fig9}. We see that there are some monthly patterns for different portfolios. Most portfolios  perform well in January and February and relatively poorly in August and September. The 
only exceptions are the Energy and the Utilities sectors  that fare well in April and May 
and poorly in September and October. This effect has been documented in the literature. See \cite{changpinegar1989}, \cite{choi2008}, and the references therein. There are many possible reasons. For example, the `January effect' might be due to that the tax-loss selling pressures  temporarily drove the security prices below their equilibrium levels in December 
and led to abnormal gains in the subsequent January when the pressures disappeared. 
Also, the study by \cite{choi2008} suggests  
that the forward looking nature of stock
prices combined with the negative economic growth in the last quarter causes the
September effect, especially in the fall season when most investors become more risk
averse and the stock prices reflect the future economic growth more than the rest of the year.

We apply the proposed method to the seasonally adjusted series in search of common factors. 
Similarly to Example 1, we also chose $m=2$ in Equation (\ref{est:m}). 
The test statistic gives $\wh r=3$, i.e. we have three estimated common factors and the rest nine  transformed series are white noises. The estimated transformation matrix $\wh\bL$ multiplied by 100 and the sample autocorrelation functions of the 12 canonical variates are displayed in the online supplement, from which we further confirm that the last nine canonical variates have 
no significant serial correlations. 

To illustrate the advantages of the proposed dimension reduction method, we compare the performance of our method with that in \cite{LamYaoBathia_Biometrika_2011} via out-of-sample 
prediction. For $h$-step ahead forecasts, we compare the observed and predicted returns 
when the models are estimated using data in the time span $[1,\tau]$ for $\tau=1002,...,1103-h$, and the $h$-step ahead forecast error is defined by 
\begin{equation}\label{ferr}
\text{FE}_h=\frac{1}{100-h+1}\sum_{\tau=1002}^{1103-h}\left(\frac{1}{\sqrt{p}}\|\wh\by_{\tau+h}-\by_{\tau+h}\|_2\right),
\end{equation}
where $p=12$ in this example. We denote GT$_1$ the proposed method without the seasonal  adjustment, GT$_2$ the proposed method with seasonal adjustment, LYB the method of \cite{LamYaoBathia_Biometrika_2011}, 
and VEC the method of applying VAR models directly to $\by_t$. 
The estimated number of common factors is $\wh r=3$ for both GT$_1$ and GT$_2$ using the canonical correlation analysis, and the number of common factors obtained by the ratio-based method in \cite{LamYaoBathia_Biometrika_2011} is $\wh r=1$. Then, we use VAR($d$) models, 
with $d$ = 1, 2, and 3, to fit the factor processes obtained by GT$_1$ and GT$_2$, 
and scalar AR($d$) models, with $d$ = 1, 2, and 3, to the 
univariate factor process obtained by LYB. For simplicity, we use AR to denote AR or VAR models, and  the $h$-step ahead forecast errors are reported in Table~\ref{Table5} for $h=1, 2$ and $3$. The smaller forecast errors are in boldface for each AR model used in the prediction and each step $h$, and similar patterns can be found for other choices of $h$. We see, from Table \ref{Table5}, the direct VAR models produce the worst predictions, which is due to over-parametrization when $p$ is large. Our methods GT$_1$ and GT$_2$ perform better than LYB. In particular, the seasonally adjusted method GT$_2$ performs slightly better than the unadjusted method GT$_1$. This result shows not only the advantages of the proposed dimension-reduction method, but also the necessity of seasonal adjustment.

\begin{table}\scriptsize{
\begin{center}
 \setlength{\abovecaptionskip}{0pt}
\setlength{\belowcaptionskip}{3pt}
 \caption{The 1-step, 2-step and 3-step ahead forecasting errors. Standard errors are given in the parentheses. GT$_1$ denotes the proposed method without the seasonal part, GT$_2$ denotes 
 the proposed method with seasonal adjustment,  ‘LYB’ is the one in \cite{LamYaoBathia_Biometrika_2011}, and VEC denotes the direct vector AR method.} \centering
          \label{Table5}
\begin{tabular}{c|ccc|c|c}
\hline 
\hline
&&GT$_1$&GT$_2$&LYB&VEC\\
\hline
$1$-step&AR(1)&{\bf 3.86} (2.07)&{\bf 3.86} (2.09)&3.87 (2.08)&3.97 (2.08)\\
&AR(2)& {\bf 3.87} (2.07)&{\bf 3.87} (2.09)&3.88 (2.05)&4.01 (2.05)\\
&AR(3)& 3.90 (2.06)&{\bf 3.89} (2.07)&3.91 (2.06)&4.05 (2.01)\\
\hline
$2$-step&AR(1)& 3.80 (2.03)&{\bf 3.79} (2.04)&3.82 (2.03)&3.82 (2.04)\\
&AR(2)& 3.80 (2.03)&{\bf 3.79} (2.02)&3.82 (2.01)&3.84 (1.97)\\
&AR(3)& 3.83 (2.02)&{\bf 3.81} (2.01)&3.85 (1.99)&3.89 (1.97)\\
\hline
$3$-step&AR(1)& 3.81 (2.03)&{\bf 3.80} (2.03)&3.83 (2.03)&3.85 (2.04)\\
&AR(2)& 3.81 (2.03)&{\bf 3.80} (2.03)&3.83 (2.03)&3.84 (2.03)\\
&AR(3)& 3.83 (2.01)&{\bf 3.82} (2.02)&3.85 (2.02)&3.91 (2.00)\\
\hline
\hline
\end{tabular}
          \end{center}}
\end{table}

\section{Concluding Remark}
In this paper, we proposed a structural-factor approach for high-dimensional time series analysis and 
demonstrated its applications with a 15-dimensional series of weekly PM$_{2.5}$ and  the monthly value-weighted returns of 12 U.S. Industrial Portfolios. For the PM$_{2.5}$ data, we do not 
consider explicitly the spatial structure of the monitoring stations. 
One can treat the proposed model as a specification for the 
dynamic dependence of the conditional mean, which can be augmented with a spatial 
covariance specification, if needed. Such an extension would be useful, especially if one is 
interested in predicting the PM$_{2.5}$ at a location not far away from the monitoring stations.  In the Industrial Portfolios data, the proposed method suggests a substantial dimension 
reduction and produces  more accurate out-of-sample forecasts.

\section*{Supporting Information}
The supplementary material contains all  technical proofs of the theorems in Section 3 and some additional Tables and Figures of the real examples of Section 4.

\section*{Acknowledgments}
 We are grateful to the  Editor, Associate Editor  and the anonymous referees for their insightful comments and suggestions that have substantially improved the presentation and quality 
 of the paper.  This research is supported in part by the Booth School of Business, University of Chicago.
 
 \section*{Data Availability Statement}
 The PM$_{2.5}$ data used in Example 1 are available in the online Supporting Information file: {\em 
 weekly-TaiwanSouth-PM25.csv}. The data used in Example 2 are available at the website cited.

\end{document}